# Models for Predicting Transonic Flutter of a Wing-Section with Sloshing in an Embedded Fuel Tank

S. Srivastava[*], M. Damodaran[**] and B.C. Khoo[#]

*National University of Singapore*
*Faculty of Engineering, 9 Engineering Drive 1, Singapore 117575*

## ABSTRACT

The present study focuses on the development, application, and comparison of three computational frameworks of varying fidelities for assessing the effects of fuel sloshing in internal fuel tanks on the aeroelastic characteristics of a wing section. The first approach uses the coupling of compressible flow solver for external aerodynamics integrated with structural solver and incompressible multiphase flow solver for fuel sloshing in the embedded fuel tank As time-domain flutter solution of these coupled solvers is computationally expensive, two approximate surrogate models to emulate sloshing flows are considered. One surrogate model utilizes a linearised approach for sloshing load computations by creating an Equivalent Mechanical System (EMS) with its parameters derived from potential flow theory. The other surrogate model aims to efficiently describe the dominant dynamic characteristics of the underlying system by employing the Radial Basis Function Neural Networks (RBF-NN) using limited CFD-based data to calibrate this model. The flutter boundaries of a wing section with and without the effects of fuel sloshing are compared. The limitation of the EMS surrogate to represent nonlinearities are reflected in this study. The RBF-NN surrogate shows remarkable agreement with the high-fidelity solution for sloshing with significantly low computational cost, thereby motivating extension to three-dimensional problems.

## I. Introduction

Sloshing of fuel in partially filled fuel tanks has been a problem of interest to researchers in the field of shipbuilding industries, vessel manufacturers, aerospace industry as well as LNG carrier automobile industry for many decades. Sloshing of fuel in an enclosed tank attached to an aircraft wing is known to affect the aeroelastic characteristics of the wing. Low aspect ratio highspeed aircrafts designed for high maneuverability such as F-16 have external stores as fuel tanks. Larger commercial aircrafts contain liquid fuels in tanks embedded inside the wing structure. Although studied to a lesser extent, there have been some important studies on the effects of sloshing on the aeroelastic motion of the wing. One of the earliest attempts on studying the effects of fuel sloshing on aircraft motion was that of Luskin and Lapin [1]. The effects of sloshing on upper subsonic and transonic flights have been recognized by the aircraft design community in Cazier et al. [2], Farhat [3], Chiu and Farhat [4], Firdous-Abadi [5] and Hall [6]. The approaches taken by these researchers vary in modeling of aerodynamics, as well as sloshing flows. Smoothed particle hydrodynamics of Monaghan [7] has been used by Banim [8] to model fuel sloshing in a wing tank, but applied a gust as the forcing function for the wing instead of aerodynamic loads. Despite the efforts of these researchers, comparably low attention has been given to the study of the effects of sloshing on aeroelastic motion in open literature.

---

[*] Graduate Student, Dept. of Mechanical Engineering; shashank.srivastava@u.nus.edu
[**] Senior Research Scientist, Temasek Laboratories; Associate Fellow AIAA; tslmura@nus.edu.sg
[#] Professor of Mechanical Engineering and Director Temasek Laboratories; tslhead@nus.edu.sg

One of the earliest approaches to model sloshing loads was the use of a suspended pendulum, tunable parameters being the mass and pendulum length. Other early approaches include set of spring-mass-damper systems based on empirical experimental data. Collectively, these representative models based on simple mechanical systems are termed as Equivalent Mechanical Systems (EMS). Later formulations based on potential flow theory based have been used to estimate parameters of EMS modeling the effects of sloshing on the structure. The linear nature of potential flow theory restricts its applications to linear domain, i.e. small amplitude motion. However, sloshing can become violent with a slight increase in motion amplitude and frequency and nonlinear phenomena such as wave-breaking and violent mixing starts taking place. High order methods such as the Volume of Fluid (VOF) method outlined in Hirt and Nicholas [9] based on Navier-Stokes equation for two incompressible and immiscible fluids capable of tracking the interface can accurately represent flow nonlinearities in enclosed sloshing flows. However, these higher order schemes are computationally expensive and are not practical for multiple runs required for problems involving iterative designing and optimization. This calls for the development of a low-cost surrogate model that can efficiently and accurately predict the dominant dynamics of coupled aeroelastic-sloshing multiphysics system which can take flow nonlinearities into account. Lucia et al. [10] and Dowell and Hall [11] provide a comprehensive overview of several reduced-order techniques such as harmonic balance, Volterra theory [12] and Proper orthogonal decomposition (POD) [13], while demonstrating their application to aeroelastic test cases. There are approaches that use linear system identification concepts to obtain a reduced-order model (ROM); however, such methods based on the state-space approach cannot accurately capture nonlinearities of the sloshing flows, large amplitude vibrations and limit cycle oscillations, which require specialized methods for nonlinear system identification.

Faller and Schreck [14] proposed a recurrent multi-layer-perceptron neural network (MLP-NN) to predict unsteady loads for aeroelastic applications and this study was subsequently followed up by Voitcu and Wong [15] and Mannarino and Mantegazza [16] leading to a systems model aeroelastic behavior of airfoils and wings based on non-recurrent MLP-NN. The dynamic loads of sloshing fluid in a fuel tank not only depends on the current state but also on the previous states and inputs since the fluid is always in a transition. In order to include dynamic memory effects, the temporal history of the excitation signal is added to the input vector of the neural network. Neural networks have been shown to offer a powerful tool in modeling nonlinear systems over a compact set rather than a small neighborhood around the dynamically linearized steady-state ROM based approach for linear systems. It provides a powerful tool for learning complex input-output mappings and has simulated many studies for the identification of dynamic systems with unknown nonlinearities as outlined in a number of studies such as Narendra and Parthsarathy [17] and Elanayar and Shin [18]. The Radial basis function neural networks (RBF-NNs) belongs to the domain of artificial neural networks, which is able to approximate any nonlinear function to an arbitrary degree of accuracy with a finite number of neurons. Originally published by Lowe and Broomhead [19], they are derived from the theory of functional approximation and conventional approximation theory. RBF-NNs have proved to a good alternative to multi-layer perceptron (MLP) networks because of their high learning rate, as detailed by Leonard and Kramer [20], and robustness to noise. The predictive capability of RBF-NN can be utilized for developing a surrogate model for predicting sloshing loads on the container due to its motion. Fuel sloshing is a transient phenomenon where previous structural state inputs, sloshing loads as well as the current structural state inputs are required for the prediction of the loads for the next time step (the future state). One widely used system identification technique is the Autoregressive technique with Exogenous inputs (ARX) as outlined in Billings [21] which assumes that the known relationship between a finite series of former inputs and previous outputs is sufficient to predict system response to subsequent inputs. Wintler [22] has used a similar technique for the prediction of aerodynamic loads using RBF-NNs.

The focus of this work is on the development, evaluation and comparison of computational model for assessing the effects of fuel sloshing in a fuel tank embedded in a wing section in transonic flow. Integrated open-source CFD solvers *SU2* [23] and *OpenFOAM* [24] are used for the solution of the coupled aeroelastic problem with sloshing in an embedded fuel tank. The *preCICE* [25] solver interface modified in Srivastava et al. [26] is used to accommodate the coupling of aeroelastic solver and multiphase sloshing solver by enabling data exchange between the solvers during runtime. The component CFD solvers march the solution in time to yield high-fidelity numerical solutions which are considered as ground truth for comparing and evaluating the predictions from other models. Two approaches for developing low-cost models for sloshing loads are then considered. The first surrogate model is an approximate EMS model in which the EMS parameters are derived by comparing the force and moment equations with those obtained from potential flow formulations for sloshing in rectangular tanks. The EMS model simulates inertial and convective components of sloshing liquids by considering linear sloshing modes. The next approach is a machine learning surrogate model based on RBF-NN to efficiently simulate the dominant static and dynamic characteristics of the sloshing tank. The dynamic memory effects are accounted by supplying the previous outputs of RBF-NN in form of loads, as well as, previous inputs in the form of structural displacement to the fuel tank. A limited set of CFD-based data from the high-fidelity computational framework is used generating ground truth data for training this surrogate. Once trained, the RBF-NN is subsequently fed with arbitrary new inputs in the form of structural motion, as well as its previous outputs to predict sloshing loads at the current time step. The sloshing loads from these two approaches are then coupled with the aeroelastic solver in the time domain in which the external transonic aerodynamics is computed using a CFD solver. The first approach using time-domain coupling of high-fidelity solvers is described in Section II. The second and third approaches using approximate surrogate models consisting of EMS and RBF-NN models for sloshing flow are described in Sections III. The flutter boundary of the NACA64A010 wing section is computed by simulating sloshing loads with surrogate models and compared with that obtained using high-fidelity CFD. Section IV discusses the results from all these approaches and compares the prediction accuracy and validity in the context of their effects on aeroelasticity of a wing-section.

## II.  Aerostructural Fuel Sloshing Model

The motion of a two degree of freedom (2-DOF) wing section embedded with a partially-filled fuel tank free to plunge and pitch. The system is represented in Fig. 1 where the airfoil motion is modeled by equivalent springs in the plunging and pitching hinged at the elastic axis. The equivalent system of airfoil embedded with a rectangular fuel tank.

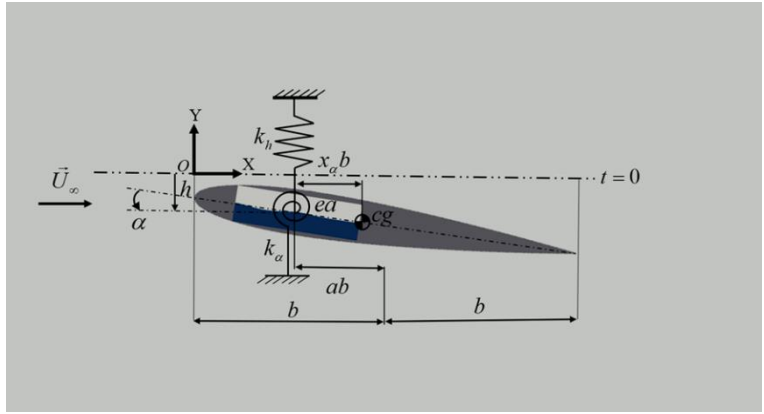

**Figure 1: A 2-DOF airfoil with a partially filled embedded fuel tank immersed in a flowfield free to plunge and pitch**

The equations of motion of this system incorporating sloshing loads is defined as

$$m_{tot}\ddot{h} + S_{\alpha,tot}\ddot{\alpha} + k_h h = -L + F_{Y,ea} \tag{1a}$$

$$S_{\alpha,tot}\ddot{h} + I_\alpha \ddot{\alpha} + k_\alpha \alpha = M + M_{OZ,ea} \tag{1b}$$

where, $m_{tot} = m + m_f$ is the combined airfoil and fluid mass, $m_f$, in the fuel tank (kg/m), $S_{\alpha,tot}$ is the static imbalance due to combined mass (kg), $I_\alpha$ is the moment of inertia about the elastic axis, $ea$, $K_h$ is the plunging spring stiffness (N/m/m), $K_\alpha$ is the pitching spring stiffness (N), $C_L$ and $C_M$ are the aerodynamics lift and moment coefficients, respectively, and $F_{Y,ea}$ and $M_{Z,ea}$ are the loads due to sloshing in the lift and moment directions, respectively. The aerodynamic and sloshing loads on the wing section are computed at each structural time step defined by $\tau_{struct} = \omega_\alpha t$. All the moments are computed about the elastic axis of the wing section in the formulation. The equations can be condensed into the non-dimensionalized matrix form defined as

$$\frac{m_{tot}}{m}\begin{bmatrix} 1 & x_\alpha \\ x_\alpha & r_\alpha^2 \end{bmatrix}\begin{Bmatrix} \ddot{\bar{h}} \\ \ddot{\alpha} \end{Bmatrix} + \begin{bmatrix} \left(\frac{\omega_h}{\omega_\alpha}\right)^2 & 0 \\ 0 & r_\alpha^2 \end{bmatrix}\begin{Bmatrix} \bar{h} \\ \alpha \end{Bmatrix} = \frac{V^{*2}}{\pi}\begin{Bmatrix} -C_L \\ 2C_M \end{Bmatrix} + \frac{1}{mb\omega_\alpha^2}\begin{Bmatrix} F_{Y,ea} \\ 2M_{Z,ea} \end{Bmatrix} \tag{2a}$$

$$[M]\{\ddot{q}\} + [K]\{q\} = \{F_{aero}\} + \{F_{slosh}\} \tag{2b}$$

where the structural parameters for the wing section used in Eqn. 2 (a) for this study are are $x_\alpha = 0.25$, $r_\alpha^2 = 0.75$, $\omega_h = 6.2831\ rad/s$, $\omega_\alpha = 6.2831\ rad/s$ and $\mu = 75$. The right-hand side of Eqn. 2 (a) - (b) contains the forcing terms evaluated from the loads due to external aerodynamic flow and internal sloshing on the wing section. The aerodynamic and sloshing loads are computed using independent solvers and are integrated in the aeroelastic equation of motion at each time step as shown in Fig 2. In the present study, the external aerodynamic loads on the wing section are computed using *SU2* and the sloshing loads on the fuel tank embedded in the wing section are computed using the *interDyMFoam* library in *OpenFOAM*.

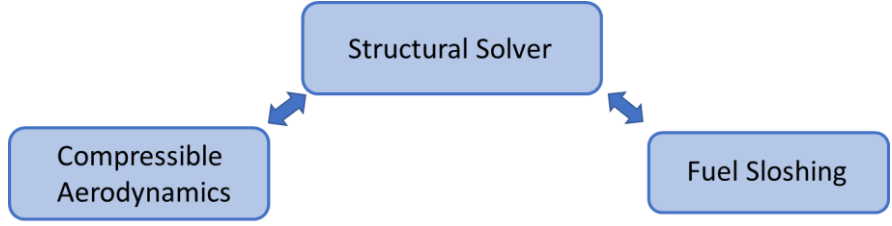

**Figure 2: Integration of aerodynamic and sloshing loads in form of force and moment with structural motion of wing section**

### A. Computation of Unsteady Aerodynamics using *SU2*

The unsteady external aerodynamic flow field is modeled using Euler equation cast into an arbitrary Lagrangian-Eulerian (ALE) formulation

$$\frac{\partial}{\partial t}\int_{\Omega(t)} \vec{U}\, d\Omega + \int_{\partial\Omega(t)} \vec{F^c}\vec{n}\, dS = 0 \qquad (3)$$

where control volume $\Omega(t)$ and its boundary $\partial\Omega(t)$ are time-dependent, $\vec{U}$ denotes the flow variables and $\vec{F^c}(U)$ is convective flux term These vectors are defined as

$$\vec{U} = \begin{bmatrix} \rho \\ \rho\vec{V} \\ \rho E \end{bmatrix} \text{ and } \vec{F^c} = \begin{bmatrix} \rho(\vec{V}-\dot{\vec{u}}_m)\vec{n} \\ \rho\vec{V}(\vec{V}-\dot{\vec{u}}_m)\vec{n} + p\vec{n} \\ \rho E(\vec{V}-\dot{\vec{u}}_m)\vec{n} + p\vec{V}\vec{n} \end{bmatrix} \qquad (4)$$

where $\rho$, $\vec{V}$, $E$, $p$ and $\vec{u}_m$ are the density, the velocity vector, fluid enthalpy, static pressure, and mesh velocity vector, respectively. The numerical method for solving Eqn. (3) and the sign convention of the aerodynamic load coefficients which is set to be consistent with that used in the compressible flow solver *SU2*, details of which are outlined in Economon et al. [23]. From the flow model, one could estimate the time variation of the aerodynamic force and moment coefficients.

### B. Computational Modeling of Sloshing in a Fuel Tank using *OpenFOAM*

The Volume of Fluid (VOF) method is used for modeling the two-immiscible media (liquid fuel and ambient gas) inside the tank where the position of the interface of the fluids is of interest. The fluid forces on the tank surfaces are computed by solving Navier-Stokes Equations in the fluid domain and the forces are calculated on the structure undergoing unidirectional forced oscillation with a single frequency. The governing equations are the unsteady, incompressible, continuity and Navier-Stokes equations. The fluid motion is described by means of the conservation of mass:

$$\nabla \cdot \vec{U}_f = 0 \qquad (5)$$

and conservation of momentum:

$$\frac{\partial \vec{U}_f}{\partial t} + (\vec{U}_f \cdot \nabla)\vec{U}_f = -\frac{1}{\rho_f}(\nabla p_f - \mu(\nabla \cdot \nabla)\vec{U}_f) + \vec{f}_B + \vec{f}_v \qquad (6)$$

where, $\vec{U}_f$ denotes the velocity of fluid relative to the tank, $p_f$ the fluid pressure, $\rho_f$ the density and $\mu$ the viscosity, respectively, $\vec{f}_B$ is the external body forces per unit mass for the liquid due to gravity and $\vec{f}_v$ represents the virtual body force per unit mass for the liquid influenced by tank motion. The sloshing problem has two phases- the incompressible fluid and the vapor regions of the tank. The VOF method is used here for volume tracking in a fixed Eulerian mesh for internal flow. In this method, a single set of momentum equations is shared by the fluid phases and the volume fraction of each fluid is tracked throughout the domain. A scalar function $f$ is used to characterize the free surface deformation, whose value is set based upon the fluid volume fraction of a cell. The volume fractions are updated by the equation written as

$$\frac{Df}{Dt} = \frac{\partial f}{\partial t} + \vec{U}_f \cdot \nabla f = 0 \tag{7}$$

After computing he volume fractions in each cell, the equivalent characteristics such as density and viscosity are defined as $\rho_{cell} = \rho_{gas} + f(\rho_{fuel} - \rho_{gas})$ and $\mu_{cell} = \mu_{gas} + f(\mu_{fuel} - \mu_{gas})$. For the present study, a 50% filled rectangular fuel tank is excited by combined pitch-plunge motion, mimicking that of a 2-DOF wing section in free aeroelastic motion. The *interDyMFoam* solver within *OpenFOAM* has been used for these simulations.

## C. Computation of Sloshing Loads

The fuel sloshing problem is solved in the time-domain and at each time-step, the sloshing forces and moments acting on the tank wall is obtained by integrating the pressure fields and shear forces along the tank walls. The pressure and shear forces are integrated over the wetted area of the fuel tank walls. The integrated forces and moments on the tank affect the aeroelastic response of the wing section. The motion of the wing section and subsequently the fuel tank, as well as the aerodynamic and sloshing loads are measured relative to the inertial frame represented by XY in Fig. 3. Freestream velocity is positive along X and the lift is measured along the Y-axis. The sloshing forces are calculated using the pressure and shear stress data on walls as follows:

$$F_{x,sl} = \int_{-h/2}^{h/2} p_E(y)w.dy - \int_{-h/2}^{h/2} p_W(y)w.dy + \int_{-l/2}^{l/2} \tau_S(x)w.dx + \int_{-l/2}^{l/2} \tau_N(x)w.dx \tag{8a}$$

$$F_{y,sl} = \int_{-l/2}^{l/2} p_N(x)w.dx - \int_{-l/2}^{l/2} p_S(x)w.dx + \int_{-h/2}^{h/2} \tau_W(y)w.dy + \int_{-h/2}^{h/2} \tau_E(y)w.dy \tag{8b}$$

and corresponding moment geometric center of the tank is :

$$M_z = \left\{ \begin{aligned} &\int_{-h/2}^{h/2} p_E(y)w(y-y_{EA})dy + \int_{-l/2}^{l/2} p_N(x)w(x-x_{EA})dx - \int_{-l/2}^{l/2} p_w(y)w(y-y_{EA})dy \\ &- \int_{-l/2}^{l/2} p_S(x)w(x-x_{EA})dx + \int_{-h/2}^{h/2} \tau_W(y)w(x-x_{EA})dy + \int_{-l/2}^{l/2} \tau_N(x)w(y-y_{EA})dx \\ &- \int_{-h/2}^{h/2} \tau_E(y)w(x-x_{EA})dy - \int_{-l/2}^{l/2} \tau_S(x)w(y-y_{EA})dx \end{aligned} \right\} \tag{8c}$$

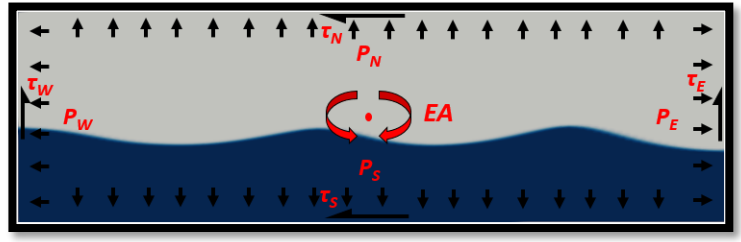

**Figure 3: Fuel tank with normal and shear loads due to sloshing fluid**

The sloshing forces are represented as the vector sum of forces on the wall, and hence there is no need to take special care of inertial forces. The projection of sloshing forces is given by,

$$F_{X,sl} = F_{x,sl} \cos\alpha + F_{y,sl} \sin\alpha \tag{9a}$$

$$F_{Y,sl} = F_{y,sl} \cos\alpha - F_{x,sl} \sin\alpha \tag{9b}$$

$$M_{OZ,sl} = M_{oz,sl} \tag{9c}$$

**D. Integration of the Component Flow Solvers:**

The external aerodynamic flow and internal sloshing flow solutions are coupled with the structural model facilitated by data transfer via interface software during the computation. The aeroelastic solver and the sloshing solver interact via transfer of forces and moments and corresponding structural displacements in the time domain at each structural time step. The solvers are coupled via an open-source coupling library for multi-physics simulations *preCICE* as shown in a workflow block diagram in Fig. 4. The *preCICE* software package includes adapters for *SU2* and *OpenFOAM* solvers which enables interaction of the solvers with the solver interface. The structural time step is governed by the aeroelastic solver i.e. *SU2* and the plunging and pitching displacements data and sloshing forces and moments are exchanged using the *preCICE* Solver Interface at each structural time step. The existing adapters are modified to facilitate coupling for the present study. The *SU2*-adapter in its native form facilitates writing of the fluid forces and moments to the solver interface which can read the structural displacements. The code for the *SU2* adapter is modified to read the integrated sloshing forces and moments and to write the structural displacements of the pitch and plunge motion on to the interface, which can be read by *OpenFOAM*. The adapter for *OpenFOAM* is modular in nature and the adapter is non-intrusive in nature. This renders higher flexibility for modification of *preCICE* adapter for the current problem. The solver used for fuel sloshing computation in internally attached fuel tank is *interDyMFoam* which is accordingly modified in the adapter. The solver interface writes displacement values to *OpenFOAM*-adapter and reads fluid forces and moments from the same.

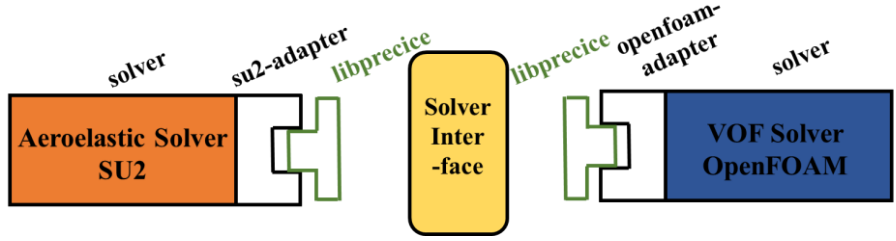

**Figure 4: Coupling of aeroelastic and sloshing solvers using a *preCICE* interface**

### III. Approximate Models for Estimation of Sloshing Loads

The present study uses two surrogate models to emulate the loads due to fuel sloshing in a partially filled rectangular fuel tank subjected to motion. These models aim to efficiently and accurately predict the dominant dynamics of sloshing fluid and reduce the computational cost of coupled solution of aero-structural-fuel sloshing system. The first surrogate model is based on a set of mass-spring system referred to as an Equivalent Mechanical System, which uses linear potential flow formulation for the solution of fluid sloshing. The second model is based on Radial Basis Function Neural Network (RBF-NN) which serves as a predictive model trained by CFD data for sloshing simulations. The derivation and formulations for both models are detailed as follows:

**A. Equivalent Mechanical Systems Characterizing Sloshing Loads**

A partially filled fuel tank is shown in Fig. 5 with the coordinate axes $x$, $y$ that is fixed to and moves with the tank, whereas the inertial coordinate system $X$, $Y$ is stationary.

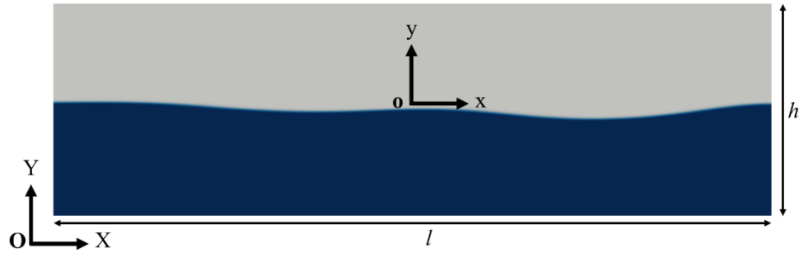

**Figure 5: Half-filled fuel tank with tank dimensions and reference axes used for derivation of Equivalent Mechanical System (EMS) parameters**

The potential flow formulation, as outlined in Ibrahim [27], is based on the assumption that the fluid in the tank is incompressible, inviscid and irrotational thereby facilitating the potential flow model which results in a general Bernoulli's Equation given by.

$$\nabla\left(\Phi_t + \frac{p}{\rho_f} - g(x - x_P)\right) = 0 \qquad (10)$$

where $\Phi$ is the velocity potential, $p$ is the pressure at any point and $\rho_f$ is the density of the sloshing liquid. The equation can be solved by applying the boundary conditions at the tank walls and the free surface. The tank of height $h$ is half-filled for the present case and the density of the gas which is present above the fluid is considered negligible compared to the liquid density, and the pressure at the free surface is considered as a static value $p_0$ at $t=0$ and the free surface is at $y = h/2$. On the free surface, one could write the dynamic boundary condition as

$$\Phi_t(x,y,t) + g\delta(x,y,t) = \frac{-p_0}{\rho_f} \quad \text{at} \quad y=h/2 \tag{11}$$

where $\delta(x,y,t)$ is free surface displacement above the initial level. The kinetic boundary condition is obtained by relating the free surface velocity with velocity potential as given by

$$\delta_t = v = \Phi_y \quad \text{at } y=h/2 \tag{12}$$

The boundary conditions can be combined to eliminate the surface perturbation $\delta$ (or velocity potential, $\Phi$) by computing the temporal derivative of $\delta$ in Eqn. (11) The resulting equation is given as follows:

$$\Phi_{tt} + g\Phi_z = 0 \quad \text{at } y=h/2 \tag{13}$$

For a forced pitching excitation of the tank with an amplitude $\psi_0$ and forcing frequency $\Omega$ about *z-axis* given by

$$\psi(t) = \psi_0 \sin(\Omega t) \tag{14}$$

with the boundary condition

$$n \cdot \nabla \Phi = i\psi_0 \Omega e^{i\Omega t} \tag{15}$$

The natural frequencies of the free surface corresponding to *sloshing modes* which are analogous to structural modes can be calculated using Eqn. 12 imposed with boundary condition given by

$$\Phi_x\big|_{x=\pm l/2} = 0, \quad \Phi_y\big|_{y=\pm h/2} = 0 \quad \text{and} \quad \Phi_y\big|_{y=0} = 0 \tag{16}$$

These natural frequencies can be shown to be as follows:

$$\omega_n^2 = (2n-1)\frac{\pi g}{l}\tanh\left((2n-1)\frac{\pi h_f}{l}\right) \tag{17}$$

where $h_f$ is the height of the free surface at $t=0$. For the present study, the fill level of the fuel tank is 50% i.e. $h_f = h/2$. The natural frequencies are dependent on the fill level, and the tank dimensions, implying that these are not influenced by the type of motion the tank configuration is subjected to. The solution of Eqn. (13) with the corresponding boundary conditions will yield the horizontal forces along the *x-axis* and moment about the *z-axis* as follows:

$$F_x = (\rho g h_f lb)\frac{\Omega^2 h_f}{g}\Psi_0 \sin(\Omega t)\left\{\frac{l^2}{12h_f^2} + \sum_{n=1}^{N}\left[F(n)\left(\frac{\Omega^2}{\omega_n^2 - \Omega^2}\right)\left(\frac{1}{2} - G(n) + \frac{g}{h\omega_n^2}\right)\right]\right\} \tag{18a}$$

$$M_z = (\rho g h_f^2 lb)\frac{\Omega^2 h_f}{g}\Psi_0 \sin(\Omega t)\left\{\frac{I_{fz}}{m_z h_f^2} + \frac{gl^2}{12h_f^3\Omega^2} + \sum_{n=1}^{N}F(n)\left[\frac{1}{2} - G(n) + \frac{g}{2h_f\omega_n^2}\right]\frac{g}{h_f\omega_n^2}\right.$$

$$\left. + \sum_{n=1}^{N}H(n)\left[\frac{1}{2} - G(n) + \frac{g}{h_f\omega_n^2}\right]^2\left(\frac{\Omega^2}{\omega_n^2 - \Omega^2}\right)\right\} \tag{18b}$$

where $F(n) = \left(8l \tanh\left((2n-1)\dfrac{\pi h_f}{l}\right)\right)\left(\pi^3 (2n-1)^3 h_f\right)^{-1}$

$G(n) = \left(2l \tanh\left((2n-1)\dfrac{\pi h_f}{2l}\right)\right)\left(\pi (2n-1) h_f\right)^{-1}$

and $H(n) = \left(8l \tanh\left((2n-1)\dfrac{\pi h_f}{a}\right)\right)\left(\pi^3 (2n-1)^3 h_f\right)^{-1}$,

$I_f$, the effective mass moment of inertia of the fluid about z-axis can be estimated as

$$I_{fz} = I_{sz}\left\{1 - \dfrac{4l^2}{l^2+h_f^2} + \dfrac{768 l^3}{\pi^3 h_f \left(l^2+h_f^2\right)} \sum_{n=1}^{N} \dfrac{\tanh\left((2n-1)\dfrac{\pi h_f}{2l}\right)}{(2n-1)^5}\right\}$$

where $I_{sz}$ is the mass moment of inertia of the solidified liquid about the z-axis at t=0. With these estimated force and moments, the tank can be represented by a configuration of mass-spring systems shown in Fig. 6, each set representative of a fluid mode. A frozen mass, $m_f$, moves in unison with the tank and represents the inertial component of sloshing, and the series of movable masses, $m_n$, represents the convective part of sloshing. Each mass is attached to the tank by a spring $K_n$, represent sloshing modes. From here, these masses will be referred to as modal masses.

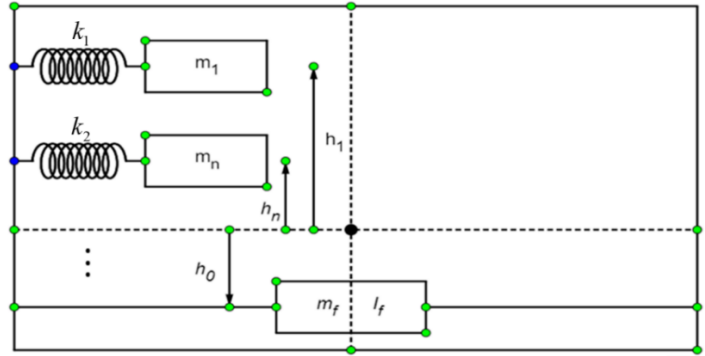

**Figure 6: Equivalent mechanical system (EMS) consisting of mass-spring system to represent sloshing loads of a partially filled fuel tank**

The modal masses, their location, and the corresponding spring constants can be computed using a series of constraints defined as follows:

- The sum of rigid and modal masses must be equal to the total fluid mass, i.e. $m_T = m_f + \sum_{n=1}^{N} m_n$

- Mass moment of inertia about z-axis which passes through the center of mass of solidified liquid, i.e.,

$$I_T = I_f + m_0 h_0^2 + \sum_{n=0}^{N} m_n h_n^2$$

- Preservation of the centre of mass, i.e., $m_0 h_0 - \sum_{n=0}^{N} m_n h_n = 0$

- Spring constants computed using modal mass and natural frequency relation, i.e., $\omega_n^2 = K_n / m_n$

This EMS configuration is subjected to the forced structural excitation about the *z-axis* defined as $\psi(t) = \psi_0 \sin(\Omega t)$. It should be noted that this structural excitation is same as that used for potential flow formulation in Eqn. 14. This is done deliberately so that the forces and moments can be directly comparable with each other. The response force and moment obtained for the EMS configuration can be shown to be

$$F_x = \Psi_0 \Omega^2 \sin(\Omega t) \left\{ m_0 h_0 + \sum_{n=1}^{\infty} m_n \left( h_n \Omega^2 + g \right) \frac{\omega_n^2}{\left( \omega_n^2 - \Omega^2 \right)} \right\} \tag{19a}$$

$$M_z = \Psi_0 \Omega^2 \sin(\Omega t) \left\{ I_0 + m_0 h_0^2 + \sum_{n=1}^{\infty} m_n \left( h_n \Omega^2 + g \right)^2 \frac{\omega_n^2}{\left( \omega_n^2 - \Omega^2 \right)} \right\} \tag{19b}$$

By comparing Eqns. 19 (a) and 19 (b) with Eqns. 18 (a) and 18 (b), the modal masses, spring constants and modal mass locations can be obtained as follows:

$$\frac{m_n}{m_f} = \frac{8}{\pi^3} \frac{\tanh\left((2n+1)\pi h_f / l\right)}{(2n+1)^3 h/l} \tag{20a}$$

$$K_n = \frac{g m_f}{h_f} \frac{8 \tanh\left((2n+1)\pi h_f / l\right)}{(2n+1)^2} \tag{20b}$$

$$\frac{h_n}{h} = \frac{1}{2} - \frac{\tanh\left((2n+1)\pi h_f / 2l\right)}{(2n+1)\pi h/2l} \tag{20c}$$

It should be noted that the modal masses, spring constants, and their respective locations are functions of tank configuration only. It is worthwhile to point out that the force and moment formulations given in Eqns. 18 (a) – (b) and Eqns. 19 (a) – (b) are linear in amplitude of the forcing motion, $\psi_0$. This is important because although the formulations and the EMS parameters are derived for forced motion of constant amplitude, the forces and moments can be easily computed for motions with varying amplitudes, such as aeroelastic motion of the wing section as long as the frequency of motion is constant. Hence different possible aeroelastic motion of the wing section such as damped response, neutral response, divergent response (flutter) and limit cycle oscillations can be addressed by the same approach.

**B. Surrogate Model for Sloshing Loads Using Neural Networks**

A Radial Basis Function Neural Network is used as a predictive surrogate model to generate sloshing loads in a rectangular fuel tank as outputs when subjected to motion, fed as inputs to the model. Sloshing fluid in a moving tank is a transient phenomenon and hence the previous structural inputs, sloshing loads as well as the current structural inputs are required for the prediction of the loads for the next time step. One widely used system identification technique is the Autoregressive technique with Exogenous inputs (ARX) outlined in Billings [21] which assumes that the known relationship between a finite series of former inputs and previous outputs is sufficient to predict system response to subsequent inputs. Such dynamical systems can be written as

$$y(t) = \bar{f}\begin{bmatrix} u(t), u(t-1), \ldots, u(t-m), \\ y(t-1), \ldots, y(t-n) \end{bmatrix} \quad (21)$$

where $y(k)$, $y(k-1)$ ... are the predicted outputs i.e. sloshing loads, for the current and previous time steps while $u(k)$, $u(k-1)$... represent the current and previous structural states. This process is shown in Fig. 7. Here, it is assumed that $\bar{f}$ is stationary for the system. Such dynamical systems which make use of sequential data and requiring some elements of 'memory' of the previous states of the systems can be modeled by Recurrent Neural Networks (RNN) for which delays must be optimized iteratively and the values of $n$ and $m$ in Eqn. 21 are determined by hit and trial method and theoretically as there are no constraints on limits of their values. For the current study, the inputs and output delay-orders are 5 and 3 respectively. The justification of these values and configuration supported by a parametric study will be outlined in Section IV.

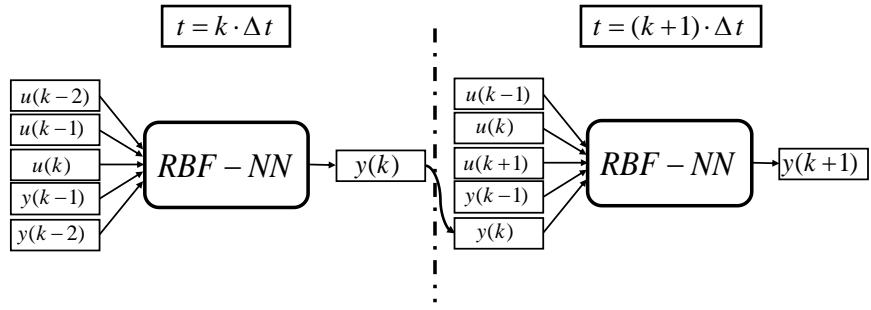

**Figure 7: ARX based network with input delay-order of 2 and output delay-order of 2**

The analysis of aeroelastic systems with sloshing in embedded fuel tanks requires coupling the CFD solvers in the time domain. The computational cost of fully coupled model is huge as fuel sloshing must be synchronized with the structure at each computational time step. Hence, it requires the development of a computationally inexpensive surrogate to model sloshing effects on the aeroelastic system. The aim of the surrogate model is to efficiently capture and describe the dominant static and dynamic characteristics of the sloshing loads on the fuel tank. Since the objective of this study is to model the sloshing forces as an unknown quantity, the approach to handle it should be same as that for a stochastic system. The sloshing forces can become highly nonlinear depending on the motion on containing structure and hence localized pressure prediction may be difficult, but it is the integrated forces and moments on the tank that will determine the motion of the container. Hence a limited CFD -based data is used to train the surrogate model which predicts integrated forces and moment on the fuel tank for subsequent structural motions fed in as inputs. Neural-network-based surrogate modeling approach is apt for this problem. Although this approach requires large quantity of training data, the computational cost of a wisely chosen training data set is still very small as compared to full-order model. The generation of the training data by structural excitation for the development of surrogate model is outlined briefly. Before an input signal is selected, it is important to identify the operating range of the system. Special care must be taken not to excite the dynamics that must not be incorporated in the model. For identification of linear systems, it is customary to apply signals consisting of sinusoids of different amplitudes or impulse inputs. However, for nonlinear model structures, it is important that all amplitudes and frequencies are represented. For the current study, the amplitude-modulated pseudo-random binary signal (APRBS), also referred to as *N*-samples-constant, is chosen for forced structural excitation to develop an input-output relation for surrogate modeling. This signal can be generated from frequently used pseudo-random binary signal (PRBS) by assigning random amplitudes to each plateau level. If *e(t)* is a white noise signal with variance, the signal defined by $u(t) = e\left(\text{int}\left[\frac{t-1}{N}\right]+1\right)$ for *t* = 1,2 … will jump to a new level at each $N^{th}$ sampling instant (*int* denotes the integer part. This signal is further modified by introducing a level change parameter with probability, = 0.5, for deciding when to change level as follows, i.e. *u(t) = u(t-1)* with probability *α* or *u(t) = e(t)* with a probability of *(1-α)*. This modification may be considered as some type of low pass filtering. Normalized white noise signal shown in Fig. 8 (a) is used for constructing the ARPBS signal which contains frequencies relevant to the system as shown in Fig. 8 (b). The APRBS signal is further modulated with level change modification is shown in Fig. 8 (c). The amplitude response to frequency as shown in Fig. 8 (d) shows that the level change parameter can be tuned to excite the frequencies of interest.

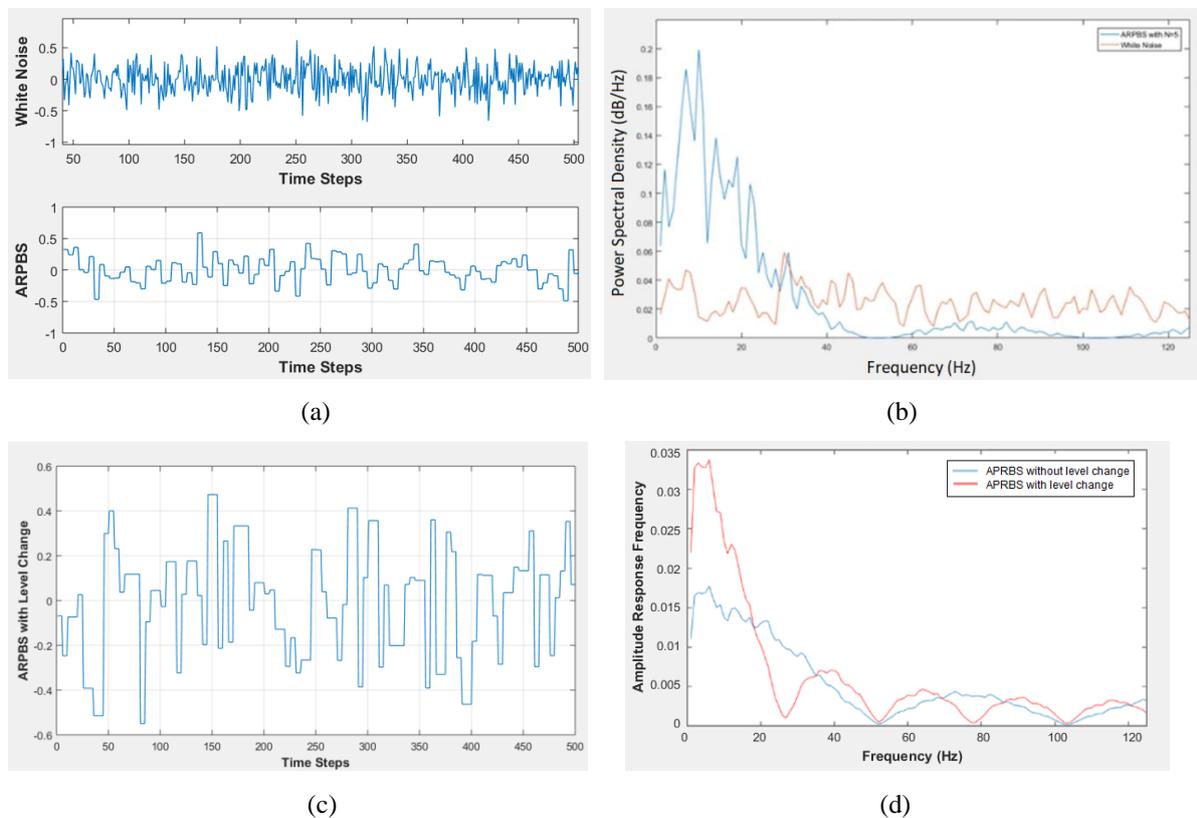

**Figure 8: (a) White noise signal used to generate APRBS and cut off higher frequencies (b) Power spectral density comparison of APRBS and white noise signals (c) APRBS with level change modulation (d) Amplitude response to excitation frequencies comparison between APRBS with and without level change parameters**

The main advantage of using the APRBS is the large spectrum of frequencies and amplitudes it offers. This property is of paramount importance for the nonlinear system identification task. Furthermore, only a short excitation time-series is needed, limiting the computational cost. Markov chain-based approach is used for training the dynamic system behavior using finite setoff input-output data samples. Using unmodulated white-noise based signal excites a larger range of frequency spectrum of the dynamics system, which is undesirable and inefficient for learning the relevant dynamics. The mathematical formulation of the Radial basis function based neural network (RBF-NN) with an output vector consisting of elements as used in the present study can be expressed as follows;

$$\tilde{y}_i = \sum_{j=0}^{M} w_{ij} \cdot \varphi_j \left( \| u - c_j \| \right) \quad \text{with} \quad \varphi_0 = 1 \tag{22}$$

where $\tilde{y}_i$ is the *i-th* element of the output vector, $W$ is a matrix containing linear weights $w_{ij}$, $u$ is the input vector and $c_j$ the center vector affiliated to the neuron *j*. The schematic representation of Eqn. (22) is shown in Fig. 9.

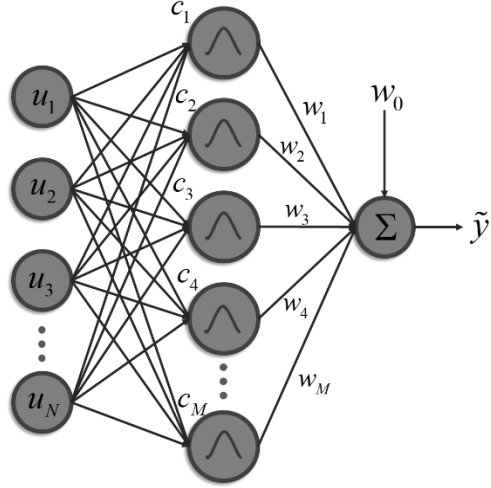

**Figure 9: Schematic representation of RBF-NN with one output y and a constant spread σ**

For the *M* basis functions in the hidden layer, the typical approach of using the Gaussian RBF in combination with the Euclidean distance norm is defined as follows:

$$\varphi_j(\| u - c_j \|) = \exp\left( -\frac{\sum_{k=1}^{N}(u_k - c_{jk})^2}{2\sigma_j^2} \right) \tag{23}$$

The spread parameter $\sigma_j$ determines the sphere of influence of neuron j and, network performance. With larger spreads, the width of the Gaussian RBF increases and therefore covers a greater regime of the input space. Recalling that only the input and output vectors are known for a given training data set, the centers, weights, and spreads must be trained to realize an RBF-NN based model.

## IV. Results and Discussions

In order to demonstrate the feasibility of the proposed RBF-NN based surrogate model and the EMS model for sloshing loads, the NACA64A010 wing section with an embedded rectangular fuel tank is considered. The external flow is governed by the Euler equations in an Arbitrary Lagrangian-Eulerian (ALE) formulation to accommodate the structural motion of the wing section. The fuel tank is considered half-filled for this study.

**A. Grid Dependence Studies for the Inviscid Aeroelastic Solver within SU2**

A NACA64A010 wing section is immersed in an external inviscid transonic flow at a Mach number of 0.70 and inclined to the flow at 3° angle of attack. The freestream temperature and pressure are 288.15 K and 101325.0 Pa respectively. The variation of computed coefficient of lift $C_L$ vs the inverse of the mesh size $N$ corresponding to the four meshes i.e. Grid 1, Grid 2, Grid 3 and Grid 4 consisting of about 2300, 4000, 8100 and 16000 elements respectively is shown in Fig. 10 (a). Fig. 10 (b) shows mesh structure in the vicinity of the wing section corresponding to Grid 3 which is used for all the flow computations in this work.

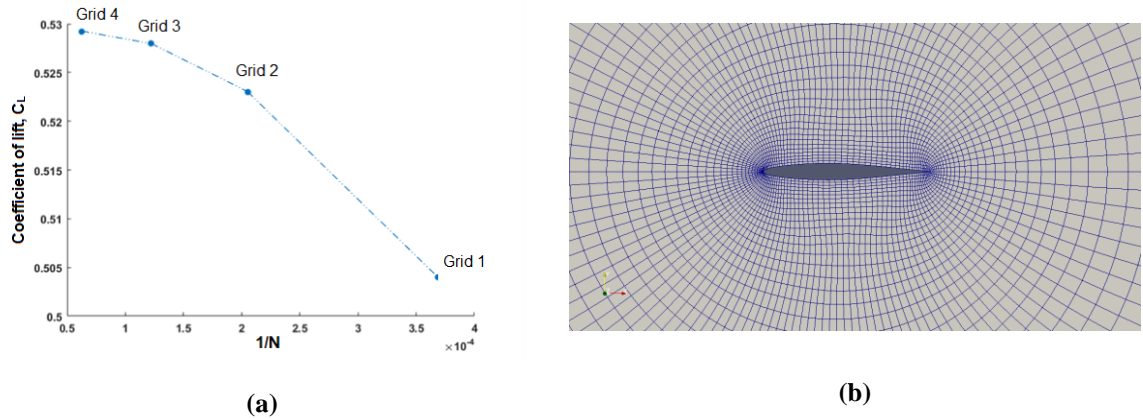

(a)          (b)

**Figure 10: (a) Variation of $C_L$ vs. 1/N and (b) Grid 3 mesh in the vicinity of the wing section**

The coefficient of lift is tracked to evaluate the order of convergence $p = \left( \ln\left[ (f_3 - f_2)(f_2 - f_1)^{-1} \right] \right)\left( \ln(r) \right)^{-1}$ with a refinement ratio, $r = 1.5$ for the grids used in the current study, i.e. Grid 2, Grid 3 and Grid 4 and where $f_1$, $f_2$, and $f_3$ are the values of $C_L$ computed on the finest, medium and coarsest mesh respectively. For Grid 3 with nearly 8100 elements, the grid convergence index estimated as in Roache [28] i.e. $GCI = F_S \left| (f_2 f_1 - 1) \right| r^{-p} - 1$ turns out to be 0.0017 for a factor of safety $F_S$ of 1.25. This lies in the acceptable range of uncertainty and hence medium grid is used to compute the aeroelastic behavior.

**B. Validation of the Compressible Inviscid Flutter Analysis of NACA64A010 Wing Section**

The compressible flow solver *SU2* is validated against numerical results presented by Alonso [29] for the case of the flutter of the NACA64A010 wing section which is forced sinusoidally in pitch for two complete cycles at a frequency close to the natural pitching frequency of the structure and then released for free motion. The three typical aeroelastic responses i.e., damping response, neutrally stable response and diverging response at different Mach number values and different values of speed index initiated by forcefully pitching the wing section at an amplitude, $\Delta\alpha$, of *1°* and frequency, $\omega_\alpha$, of *100 rad/sec*, have been replicated in the present work as shown in Fig. 11 (a) – (c).

Fig. 12 compares the computed flutter boundary at four Mach numbers using time-domain solver *SU2*. At each Mach number, the flutter speed index is fixed and the wing section is forcibly pitched for cycles with *1°* amplitude at its natural frequency, i.e. $\omega_\alpha$, of *100 rad/sec*.

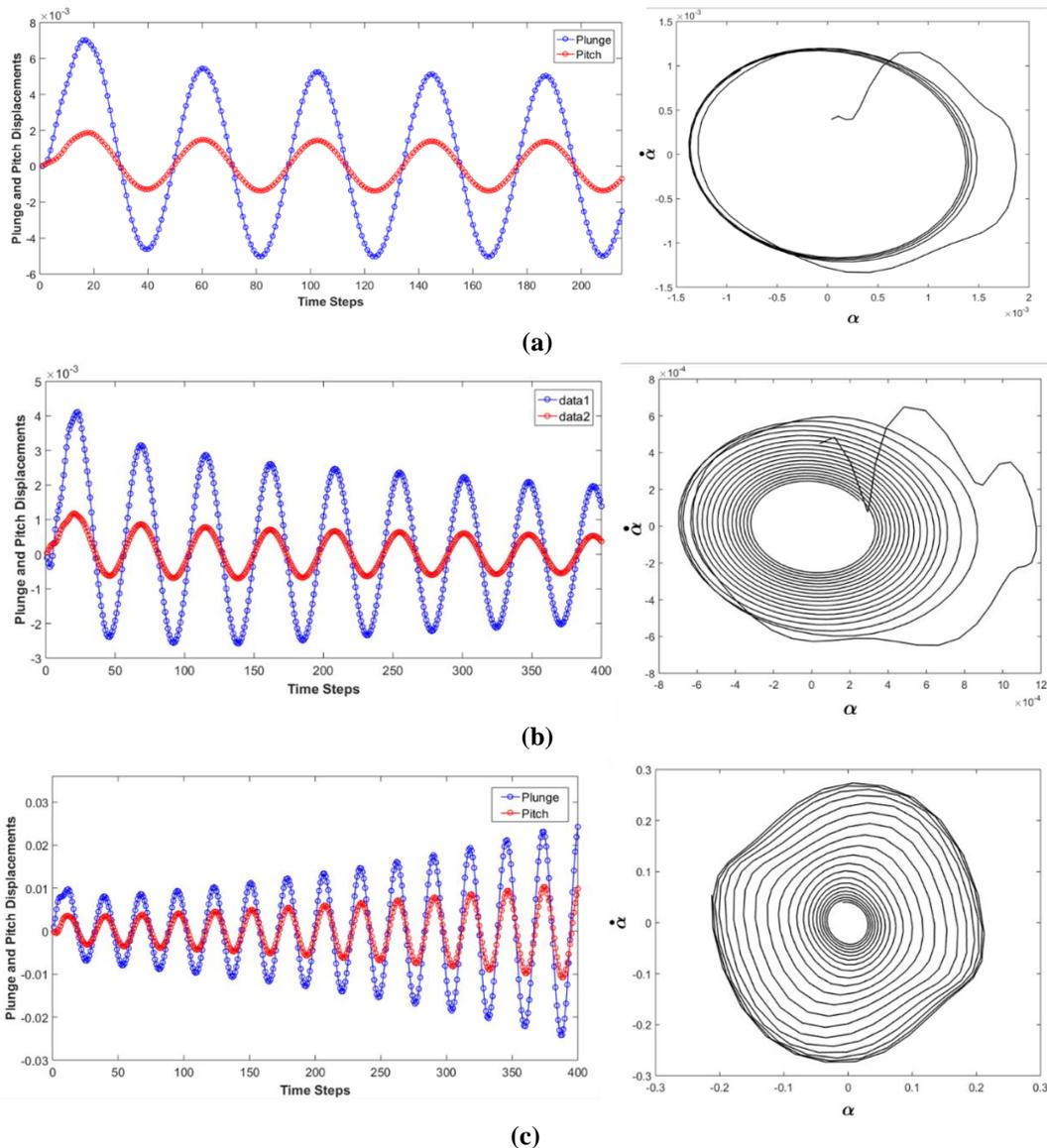

**Figure 11: Free aeroelastic motion with (a) neutral response at $M_\infty$ of 0.825 and $V_f$ of 0.612, (b) damped response at $M_\infty$ of 0.85 and $V_f$ of 0.439, and (c) diverging (flutter) response at $M_\infty$ of 0.875 and $V_f$ of 1.420**

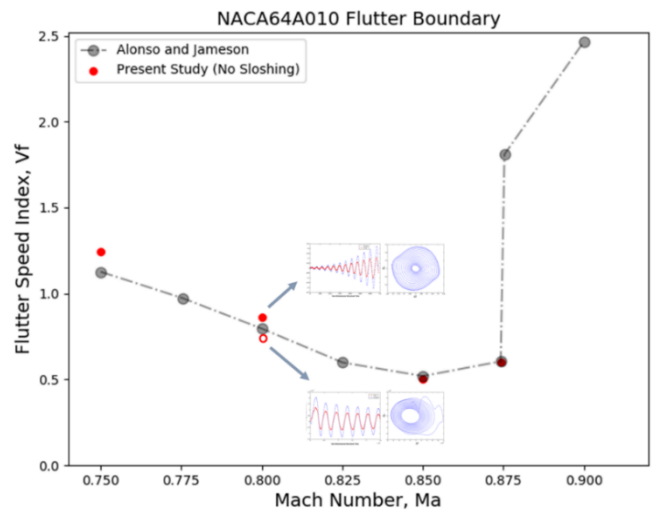

**Figure 12: Flutter boundary computation and comparison with Alonso and Jameson [29]**

Forced pitching the airfoil about the elastic axis perturbs the flowfield in the computation domain around the airfoil, which is used to initiate the aeroelastic computations. The resulting motion is identified for damped, neutral or diverging response as shown in Fig. 11(a), (b) and (c), respectively. The pitch and plunge motion and corresponding phase plot are used to identify the onset of flutter. In Fig. 10, for a Mach number, $M_\infty$ of 0.80, the flutter speed index is varied starting from flutter speed index, $V_f$ of 0.78, until 0.86, where the onset of flutter is observed. The pitch and plunge plots and phase plots are shown for damping aeroelastic response at a flutter speed index, $V_f$ of 0.78 and flutter response at $V_f$ of 0.86. The flutter boundary of the NACA64A010 wing section is verified at four points spread across the transonic range. These points are later used for comparison of the effect of sloshing on the flutter boundary. The obtained flutter boundary shows a good correlation with Alonso and Jameson's [29] results. The slight difference is attributed to the difference in mesh used for the present study and published literature and choice of flutter speed index used for computing the flutter boundary.

**C. Prediction of Unsteady Sloshing Force Using Surrogate Model and Validation Using CFD**

The RBF-NN architecture for prediction of sloshing loads in the fuel tank with an input delay of 5 and an output delay of 3 is used for predicting the sloshing loads. This configuration is established by parametrically varying the input-delay and output-delay and recording the prediction error of the RBF-NN. The prediction error is defined by the mean percentage error at all prediction points in the time series. Fig. 13 (a) shows the variation of mean prediction error with the input-delay of the RBF-NN. Similarly, Fig. 13 (b) shows the variation of mean prediction error of the RBF-NN with output delay. The architecture of the RBF-NN is converged upon an input-delay of 5 and output-delay of 3 since minimum prediction error is obtained with this configuration.

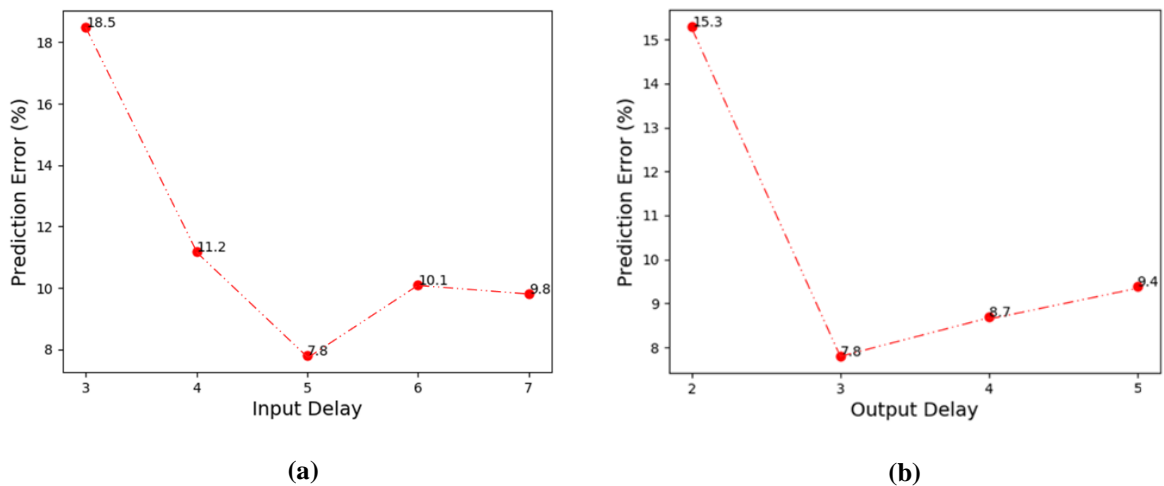

(a)    (b)

**Figure 13: (a) The structural input provided as pitching angle to fuel tank, (b) comparison of forces on east wall using CFD output and surrogate**

The RBF-NN is trained using sloshing loads response of the fuel tank when excited by plunging and pitching motion. The *N*-sample constant based signal is used as excitation inputs as shown in Fig. 14 (a), which are designed to excite the tank motion near-natural pitching frequency of the wing section i.e. $\omega_\alpha$ of 100 *rad/s*ec, as described in Section III (B). The amplitude of plunging and pitching motion imparted to the fuel tank corresponds to that of the NACA64A010 wing section in free motion. The average amplitude of five-cycles of free harmonic motion corresponding to Fig. 11 (a) is used to scale the structural motion imparted to the fuel tank for the collection of sloshing loads response for training the RBF-NN. Other forms of signals such as chirping signal **[30]** can also be used for this purpose. The computed sloshing forces and moments on the tank wall from CFD data corresponding to the structural excitation input are divided into training data and testing data. About 90% of these data points are used in the training of the neural network, i.e. calculation of weights and centers of the RBF-NN. The remaining 10% data are used for validating the RBF-NN predictions. The prediction of the unsteady sloshing loads in the tank from the RBF-NN surrogate is compared with the corresponding prediction using CFD in Fig. 14 (b) – (c) and the mean prediction error is found to be less than 8%. Fig. 14 (d) – (e) shows a zoomed-in view of comparison of CFD data and RBF-NN predictions along with percentage error for each prediction point corresponding to time steps 901 to 1000.

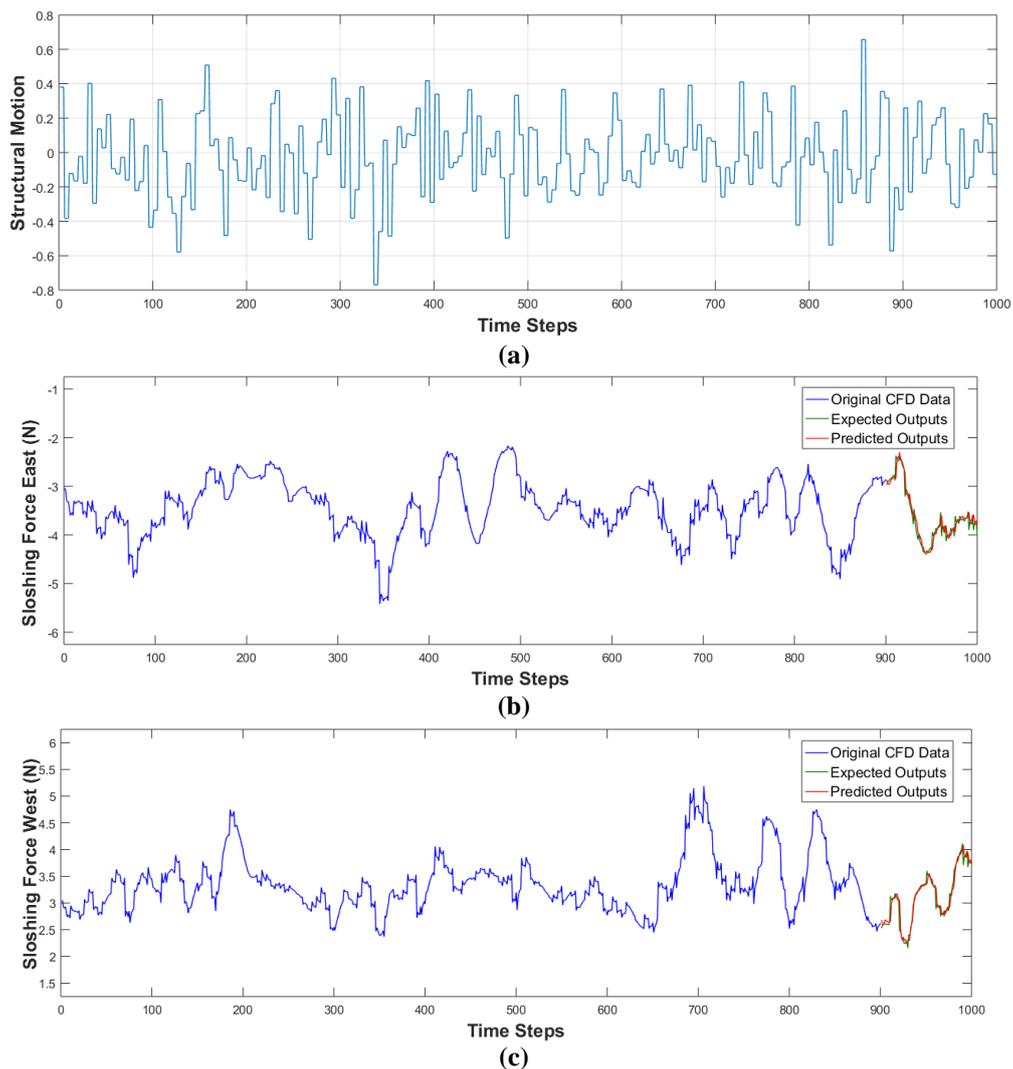

**Figure 14: (a) The structural input provided to fuel tank, (b) comparison of forces on east wall using CFD output and surrogate model prediction, and (c) comparison of forces on west wall using CFD output and surrogate model prediction (Continued)**

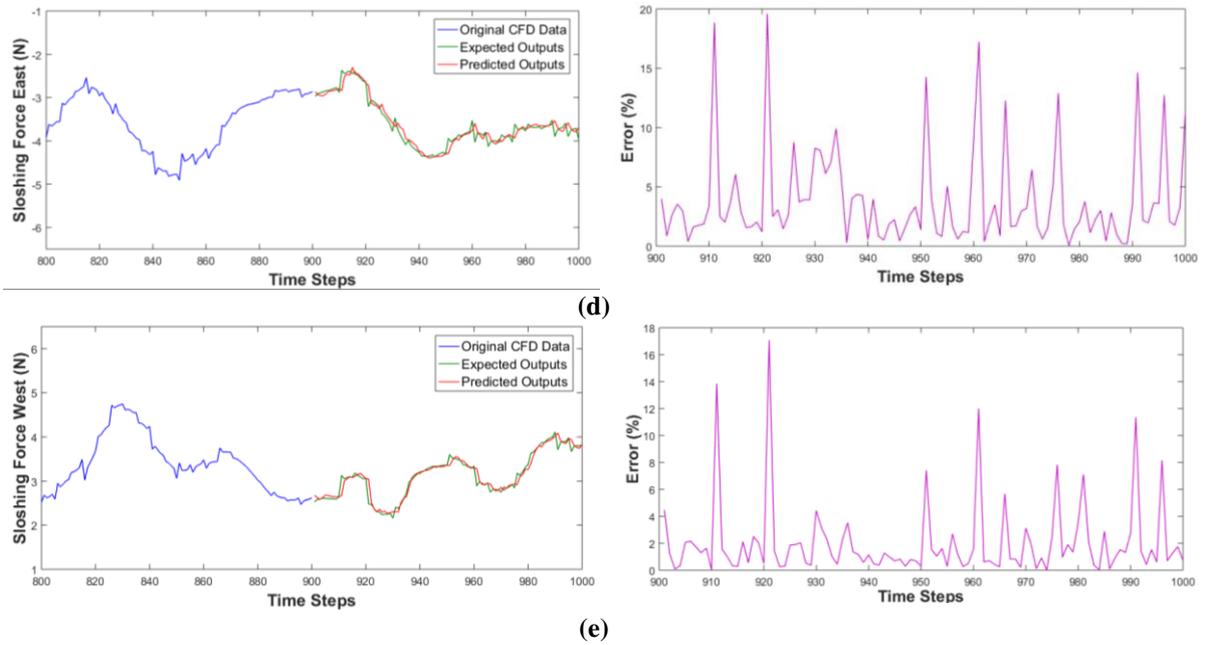

**Figure 14:** (d) zoomed-in comparison of forces on east wall and corresponding errors, and (e) zoomed-in comparison of forces on west wall and corresponding errors

The structural motion of the half-filled tank is given as a pitching excitation using the APRBS signal shown in Fig. 14 (a) for validating the predictions from the RBF-NN surrogate model. The corresponding response of the forces on the lateral walls of the tank from CFD simulation are shown in Fig. 14 (b) and Fig. 14 (c) respectively (in blue line). The forces predicted from the RBF-NN are shown in red lines on the same plots with CFD forced shown in light blue for the same structural motion. It can be seen that the force predictions from the RBF-NN are in good agreement with those from CFD predictions.

### D. Wing Section Flutter Analysis Using High Fidelity Computational Framework

The high-fidelity computational framework for modeling the behavior of the aero-structural system outlined in Section III is used to develop the flutter boundary of the aero-structural-fuel tank sloshing system. The fill level of the tank is kept at 50% and the flow field is initialized with a quasi-periodic flowfield corresponding to the wing section subjected to a pitching excitation at natural frequency and forced excitation amplitude at the given freestream transonic Mach number. After a few cycles of forced motion, the wing section is set to move freely directed by the ambient pressure field. The free motion of the wing section is captured for the first five cycles and the aeroelastic motion is determined thereafter. The computed Mach contours of the external flowfield at selected instants of time during the unsteady motion starting from the initial state are shown in Fig. 15. The steady-state flowfield used to initialize and evolve the forced pitching motion is shown in Fig. 13 (a). Similarly, the flowfield obtained at the end of the forced motion shown in Fig. 15 (b) is used as initial flow conditions for initiating the free motion of the wing section. This process not only provides an initial perturbation to the wing section but also speeds up the computations. Finally, Fig. 15 (c) and Fig. 15 (d) shows the flowfield around the wing section during its free motion, which in this case is diverging motion in plunge and pitch. Similarly, the fuel tank is forcefully pitched for 2 cycles and the flowfield is initialized for free motion of tank attached with wing section. The initial state of fluid in tank which has 50% fill level is shown in Fig. 16 (a). The perturbed fuel-vapor interface location after two forced pitching cycles is shown in Fig. 16 (b). This is followed by the free motion of the fuel tank (rigidly attached to and moving with the wing section) and fuel volume fraction is shown for two instances in the 5th cycle of motion in Fig. 16 (c) - (d).

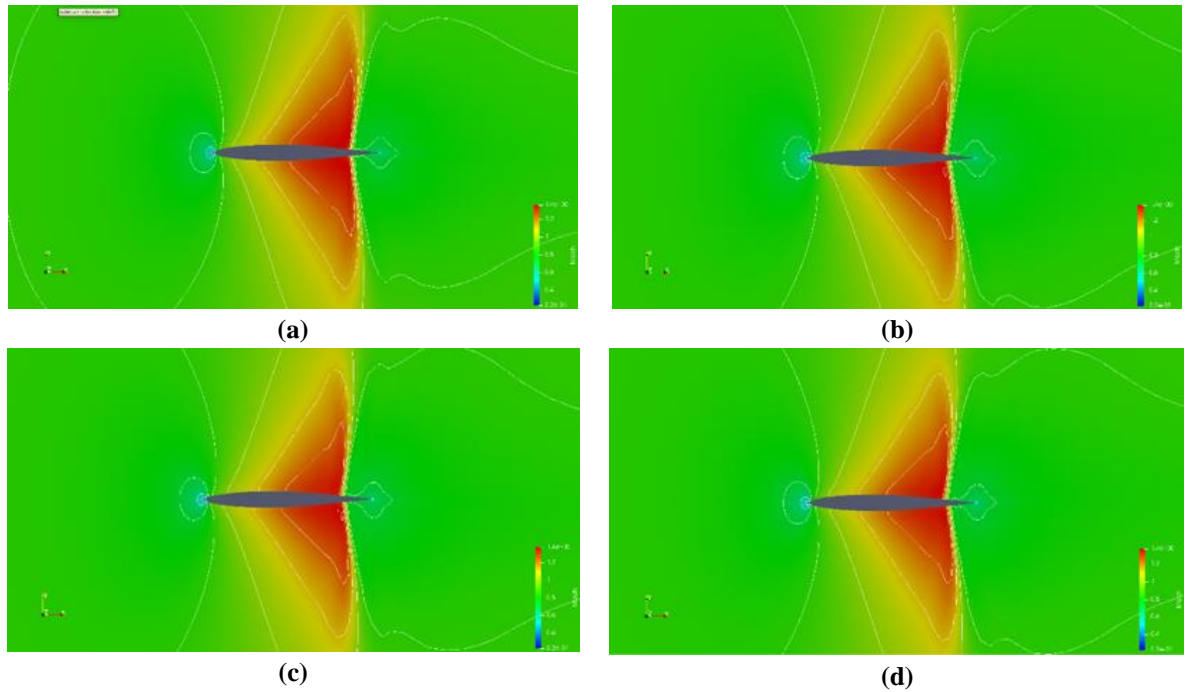

**Figure 15: (a) Computed Mach contours for steady state flow past NACA64A010 wing section at $M_\infty = 0.80$ and $V_f = 0.70$, (b) Unsteady flowfield at the end of 2 forced pitching cycles, (c) Wing Section in $\pi/4$ position in the 5th free cycle going into flutter, and (d) Wing Section at the end of 5th free cycle**

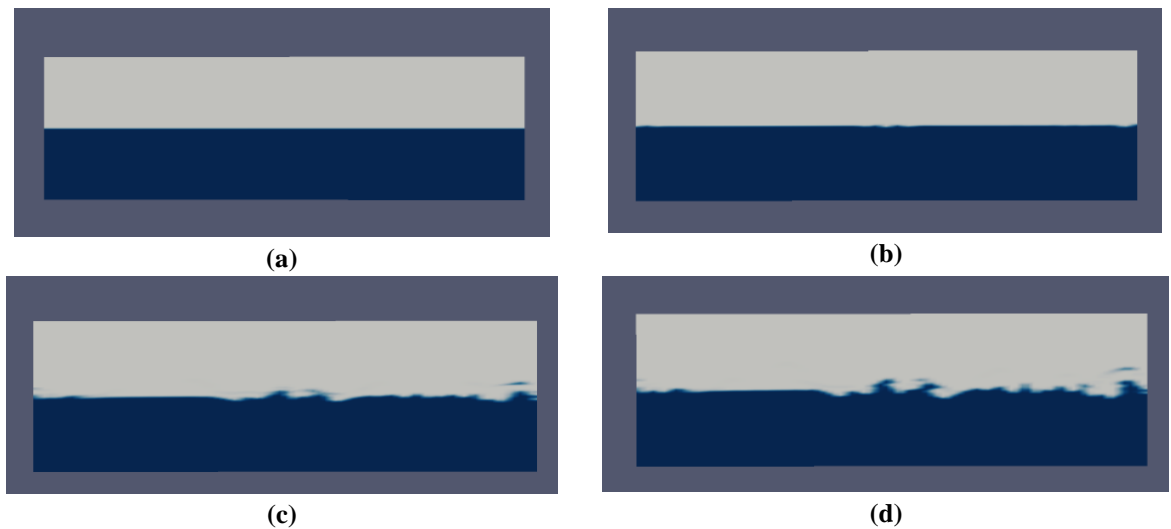

**Figure 16: (a) Volume fraction of fuel in rectangular container (50% fill level), (b) Free surface at the end of 2 forced pitching cycles, and (c) Tank at $\pi/4$ position in the 5th free cycle, and (d) Tank at the end of $5^{th}$ free cycle**

The surrogate model is trained by providing the motion of the wing section to the fuel tank in the form of a combined pitching and plunging motion. The APRBS signal is used as the input to the wing section with the half-filled tank and the corresponding sloshing loads in terms of lateral and vertical forces and moment about the pitching axis form the target outputs. The predicted sloshing forces and moments from the RBF-NN based surrogate model for the combined pitching and plunging motion of the fuel tank are used to compute the aeroelastic response of the wing section due to external aerodynamic loads and sloshing loads quickly. Flutter boundary points are computed by fixing the Mach number and incrementally increasing the flutter speed index. The aeroelastic response is analyzed in time domain and the first sign of divergence is marked as the onset of flutter. Fig. 17 compares the flutter boundary predicted using the high fidelity CFD based computational framework and the RBF-NN model considering the effects of fuel sloshing in the embedded fuel tanks with and without sloshing effects.

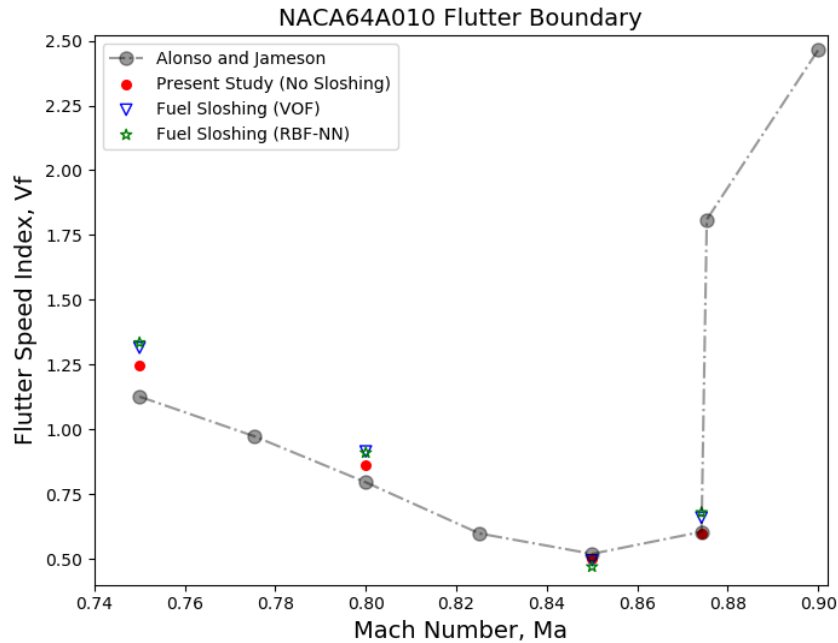

**Figure 17: Flutter boundary of NACA64A010 wing section with and without fuel sloshing effects modeled by CFD and RBF-NN based surrogate model**

The primary focus of this work is to develop a framework to study the effects of sloshing fuel in an internal tank on the flutter characteristics of aeroelastic systems. Keeping in mind the nonlinear nature of external aerodynamics as well as sloshing of fluid, two approaches of different fidelities are adopted to simulate the effects of internal sloshing in time domain. The aerodynamics, however, has been computed using CFD in time domain for all cases. The surrogate model accurately predicts the sloshing loads acting on the tank walls when subjected to the *N*-sample constant inputs for the structural motion of the tank. The flutter boundary is computed for the wing section with and without the sloshing forces. The flutter boundary is modified by the sloshing effects of fluid inside the fuel tank. These results correspond to a particular fill level (50%) in the tank, the density and viscosity of the fluid is also fixed as well as the tank size and geometry. Each of these factors will affect the flutter boundary and at this stage, their sensitivities to flutter computation are not known. The present framework can be used to study each factor individually. The RBF-NN based surrogate model for multiphase sloshing flows provides a basis for fast prediction of the flutter boundary and the onset of flutter.

**E. Flutter Boundary of the NACA64A010 Wing Section with EMS Model for Sloshing**

The fundamental problems of sloshing require approximation of hydrodynamic pressure distribution, integrated pressures to compute forces and moments generated, location of free surface and response frequencies of the bulk liquid. Broadly, there two components of hydrodynamic pressure; an inertial component as a direct consequence of container acceleration, and a convective component representing free-surface liquid motions. These are modeled using a configuration of mass-spring based Equivalent Mechanical System (EMS) to represent *sloshing modes* as shown in Fig. 6 where each set of mass and spring represents a sloshing mode. Theoretically, there are an infinite number of modes and corresponding natural frequencies. However, the lowest few modes are likely to be excited and contribute to the global response of the fluid. These modes are computed using linear potential flow theory. This can be verified by computing the EMS parameters of a partially-filled fuel tank and the sloshing loads by cumulatively adding higher sloshing modes. The tank width is 0.5 m, height is 0.15 m and the water level is 0.075 m. The parameters are computed for the first 8 sloshing modes. The computed modal masses are normalized to 1 in order to observe the contribution of each modal mass to force on the wall. The normalized modal masses, spring constants, and natural frequencies are shown in Fig. 18 (a) – (c).

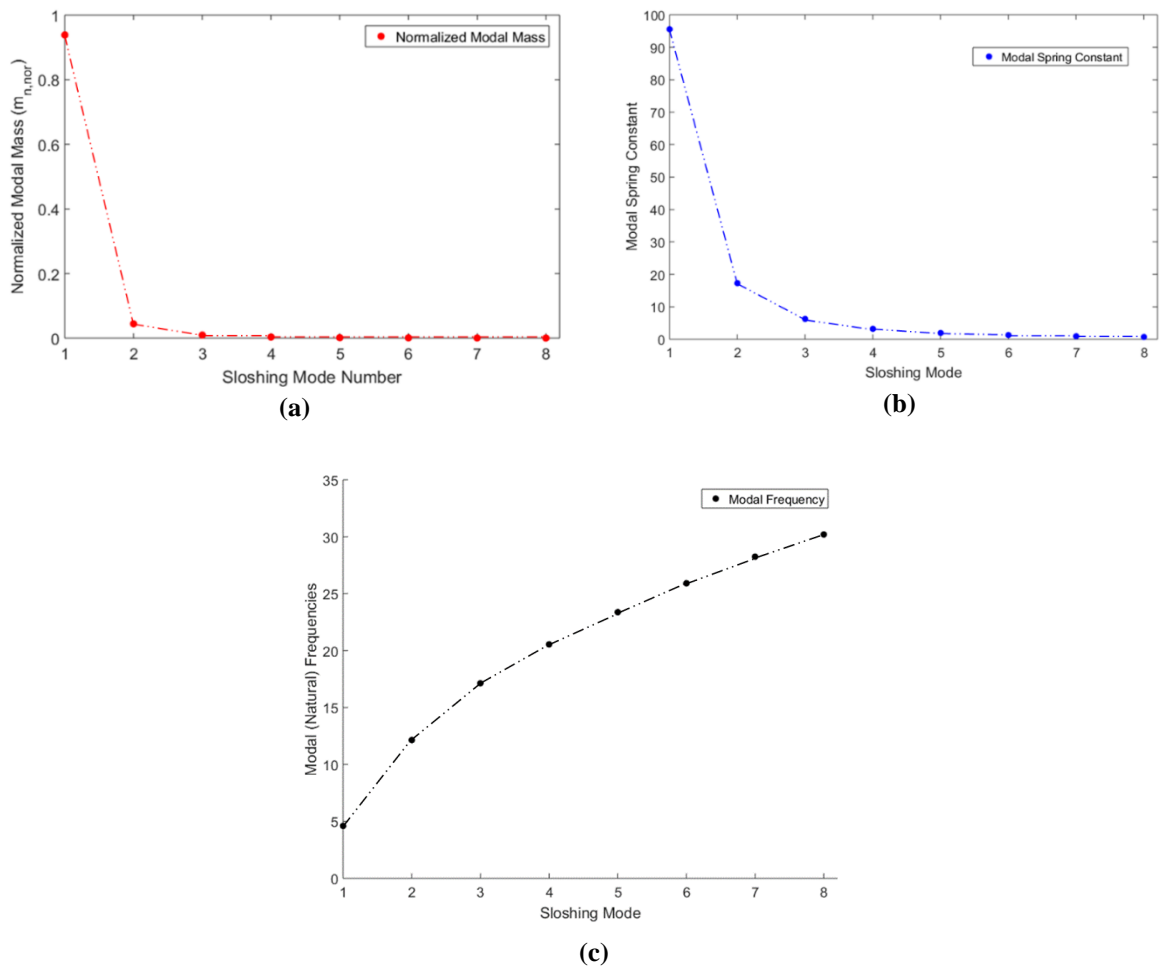

**Figure 18: (a) Normalized sloshing modal mass for first 8 sloshing modes, (b) Modal spring constants, and (c) Modal (natural) frequencies of sloshing modes**

The variation of cumulative forces on the tank walls due to sloshing with subsequent number of modes of sloshing is shown in Fig. 19. It can be seen that the first few modes play a dominant part in computing the sloshing forces and moments. This is confirmed by plotting the cumulative modal forces with increasing number of modes. The force amplitude saturates with subsequent addition of sloshing modes. The addition of 4$^{th}$ mode increases the force amplitude by just 1.14%. For the present study, only the first three sloshing modes are chosen for EMS model.

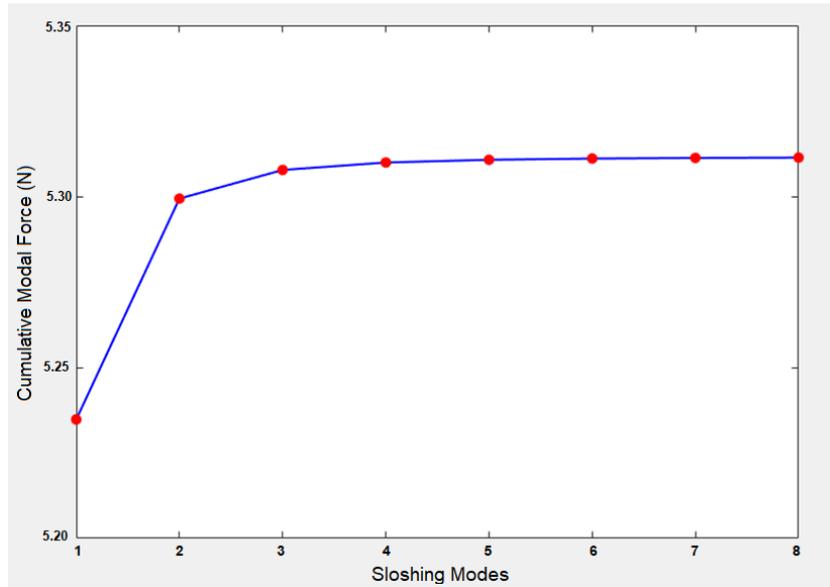

**Figure 19: Cumulative modal forces with increasing number of sloshing modes**

As discussed earlier, the sloshing loads vary linearly with forcing amplitudes. However, this holds true when the excitation frequencies are not close to the natural frequencies of the system. Near the natural frequencies, the liquid motion exhibits a nonlinear behavior in terms of sudden spike in motion amplitude, chaotic motion of the fluid, wave-breaking and nonlinear mode interactions. Fig. 20 shows the force amplitude response to forcing frequency of excitation of a partially filled fuel tank based on Eqn. 18 (a). The force response spikes when forcing frequency nears the natural sloshing frequencies of the fuel tank marking the limitation of the linear model to predict sloshing response.

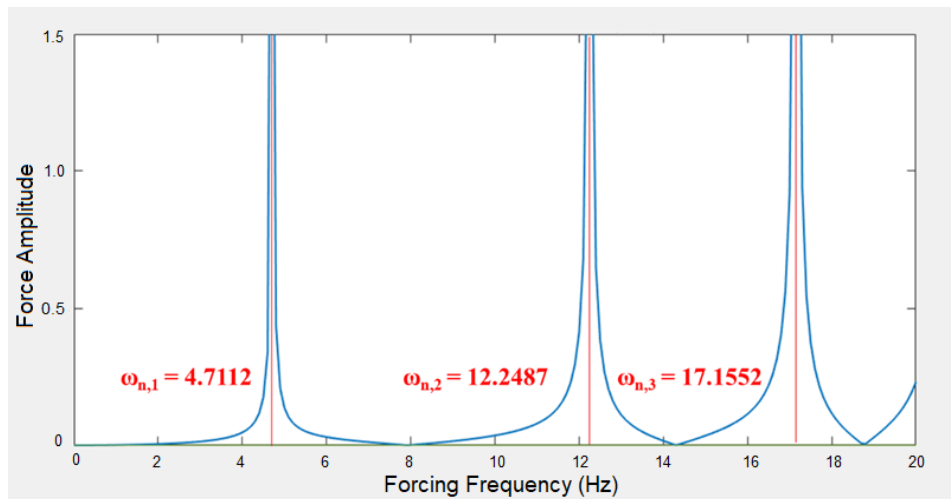

**Figure 20: Force amplitude response to forcing frequency**

The flutter boundary of the NACA64A010 wing section is now computed with the EMS model for sloshing as outlined in Section III. Fig. 21 compares the flutter boundary for the NACA 64A010 wing section with sloshing inside the embedded fuel tank estimated using the EMS model for sloshing loads with the flutter boundaries predicted using the high fidelity CFD based computational framework and the RBF-NN based surrogate model.

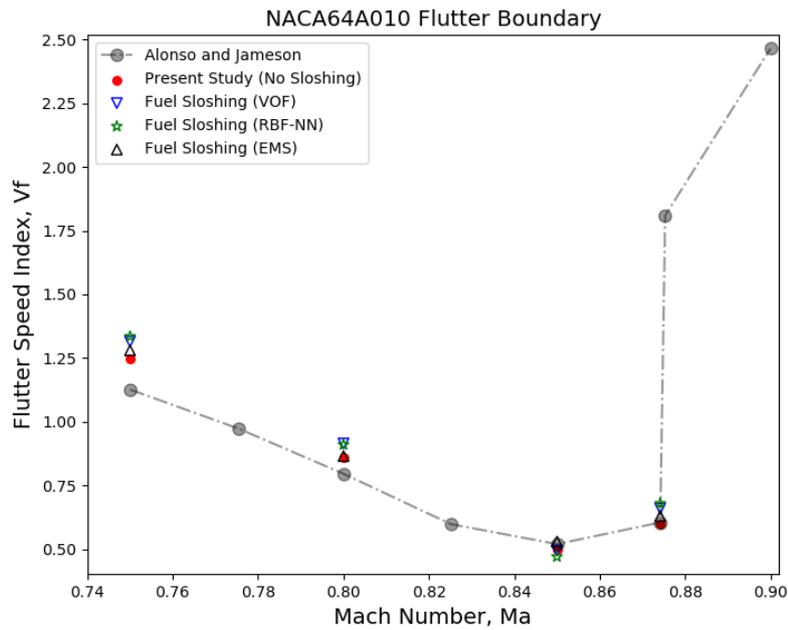

**Figure 21: Flutter boundary of NACA64A010 Wing Section with and without the effects of fuel sloshing**

The flutter boundary obtained from the EMS model for sloshing is slightly deviated from the flutter boundary obtained using high-fidelity model. This can be attributed to many factors such as absence on nonlinear effects in sloshing and near-natural forcing frequencies. It is worthwhile to compare the nature of forces obtained from CFD and the EMS model based on potential flow due to pitching motion of the tank attached to the wing section at its natural pitching frequency, i.e. 100 rad/s. The amplitude of the pitching angle is set as 1° and the resulting forces in the lateral direction are tracked as a time series. The forces due to sloshing are estimated at every 0.02 s physical time for a total time of $T$ of 4 s. A comparison of lateral forces from sloshing computed using CFD and the potential flow formulation for pitching motion is shown in Fig. 22 which shows that the sloshing forces at high frequencies are highly nonlinear which the linear potential theory cannot predict correctly.

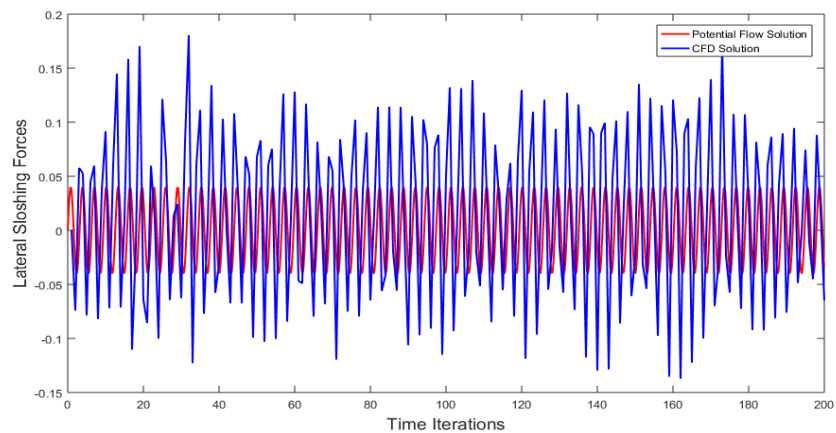

**Figure 22: Lateral forces acting on the tank walls as a result of fuel sloshing due to pitching motion of the fuel tank at f = 15.91 Hz using CFD and potential flow solution**

Since the correlation between the sloshing loads obtained from high-fidelity CFD and EMS model is not good, it is worthwhile to analyze the limits of the comparability of EMS model with CFD. First, the linear variation of sloshing loads with amplitude is verified. This is important because EMS models can only be used for aeroelastic motions if the linear relation holds. Fig. 23 shows the variation of the lateral sloshing forces on the tank wall compared for forced pitching motion of the tank for different values of pitching amplitude in the range of 1° - 8° at a constant frequency.

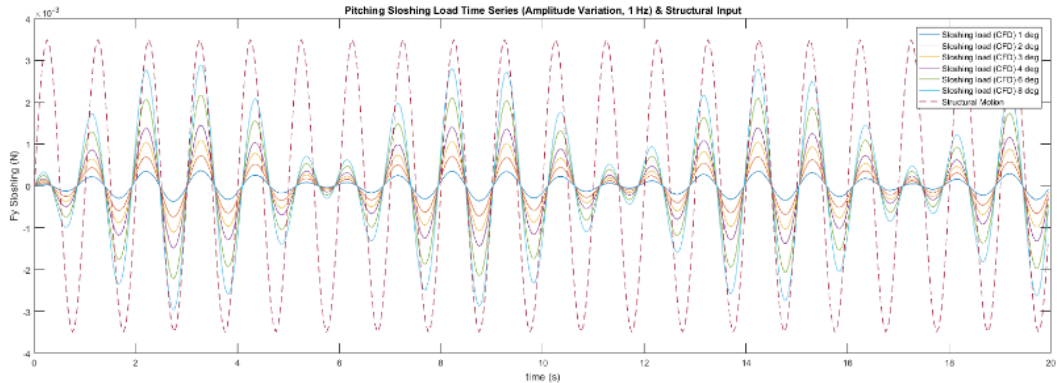

**Figure 23: Lateral sloshing forces on tank walls for forced pitching with increasing amplitude**

From Fig. 23 it can be seen that the sloshing loads obtained from CFD are linear with forcing amplitude. However, since the aeroelastic motion is a combination of plunging and pitching motion, a parametric study with amplitude variation is done for combined motion. Fig. 24 (a) – (d) shows the variation of the vertical sloshing force, lateral sloshing force and moment about the pitching direction on the fuel tank walls and the corresponding FFT plot for different values of the motion amplitudes of plunging and pitching for a fixed frequency of 3.14 rad/sec.

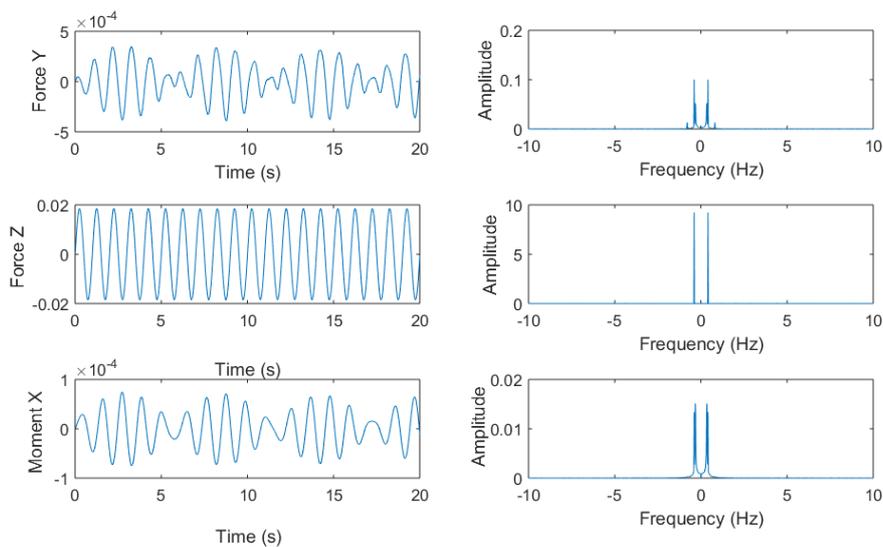

(a)

**Figure 24: Sloshing forces and moment time-history obtained from CFD and Fourier analysis for combined pitch-plunge motion with ω=3.14 rad/sec and (a) α = 1° , h = 2.5% (Continued)**

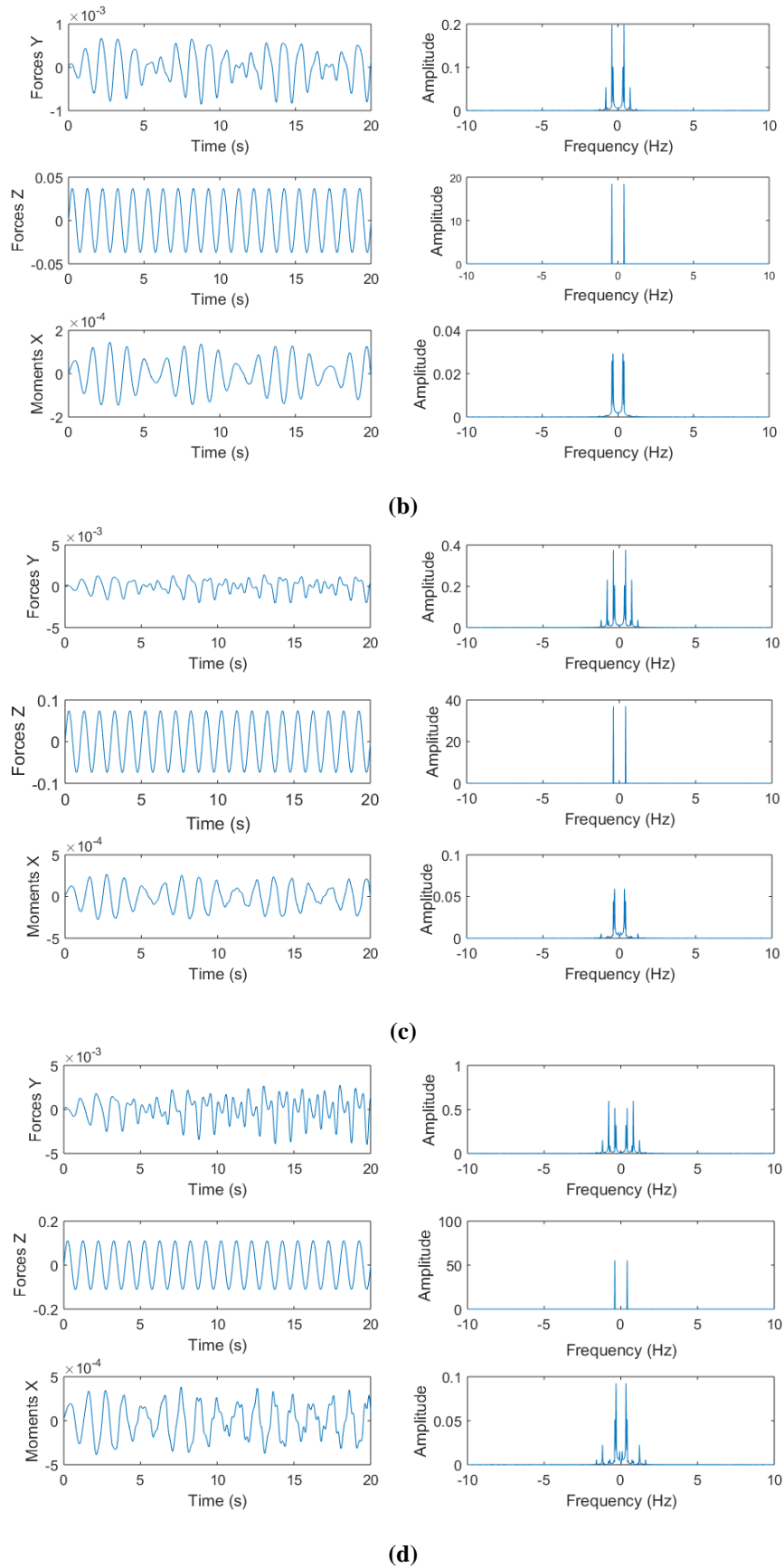

**Figure 24:** Sloshing forces and moment time-history obtained from CFD and Fourier analysis for combined pitch-plunge motion with ω=3.14 rad/sec and (b) α = 2° , h = 5.0%, (c) α = 4° , h = 10.0%, and (d) α = 6° , h = 15.0% of tank height

Although the individual force or moment profile may not be important, the appearance of fewer peaks (maximum of 2) in Fig. 24 (d) confirms that the sloshing loads obtained from CFD can be represented by a few fundamental modes, and thereby supporting the validity of the comparison of CFD and EMS for amplitude variation. Similarly, a parametric study of sloshing loads response using CFD with varying frequency for a fixed amplitude of motion of plunge and pitch is done. The variation of vertical and lateral forces and moment in the pitching direction to the tank wall is plotted along with their respective FFTs in Fig. 25 (a) – (d).

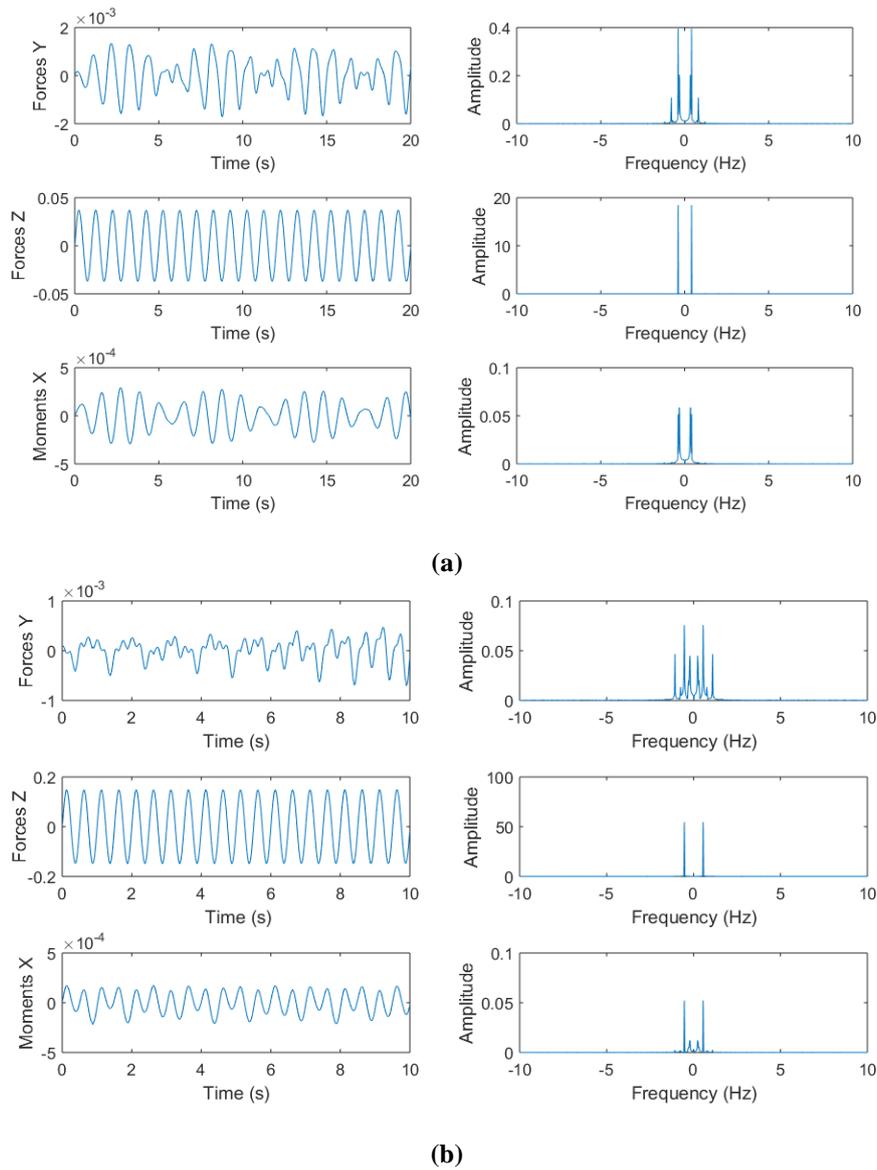

**Figure 25: Sloshing forces and moment time-history obtained from CFD and Fourier analysis for combined pitch-plunge motion with plunging amplitude h = 5.0% of tank height, pitching amplitude α = 4° and forcing frequency of (a) ω=3.14 rad/sec, (b) ω=6.28 rad/sec (Continued)**

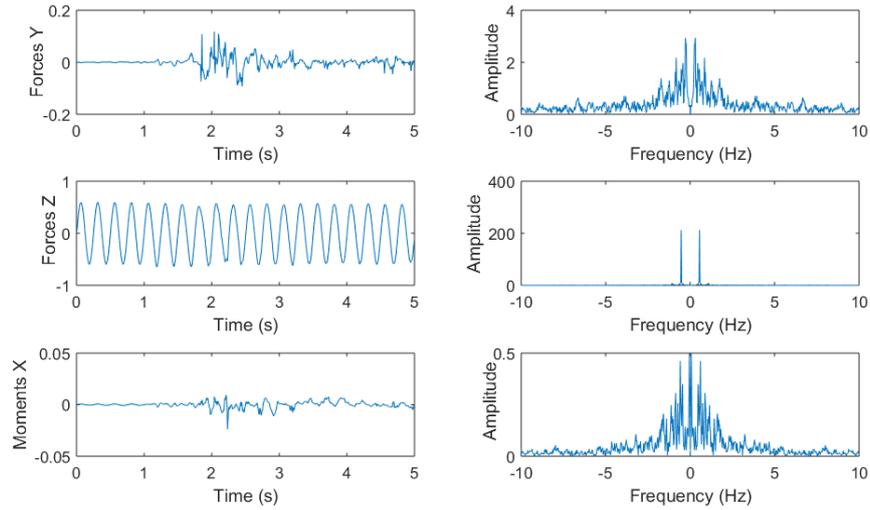

(c)

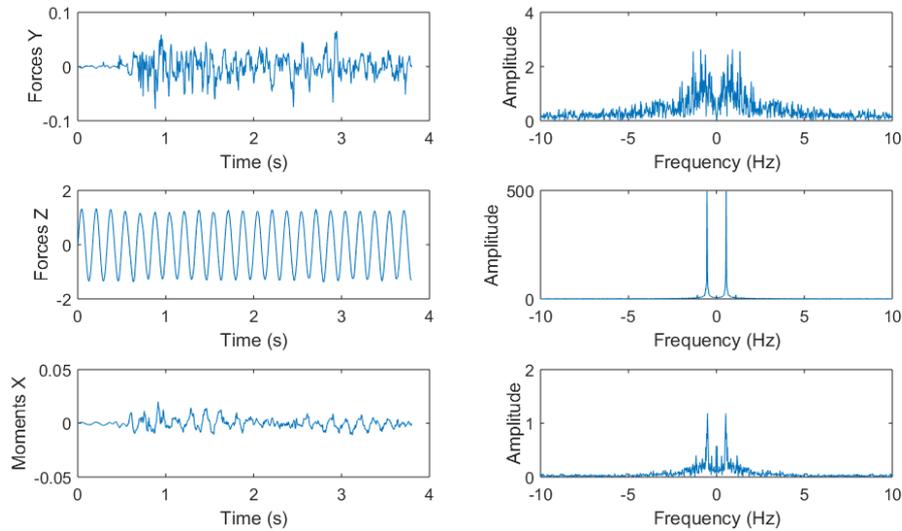

(d)

**Figure 25: Sloshing forces and moment time-history obtained from CFD and Fourier analysis for combined pitch-plunge motion with plunging amplitude h = 5.0% of tank height, pitching amplitude α = 4° and forcing frequency of (c) ω=12.56 rad/sec, and (d) ω=25.12 rad/sec**

From Fig. 25 (a) – (d) it can be inferred that the sloshing loads obtained from CFD are sensitive to excitation frequency the response becomes more and more nonlinear with higher frequencies. More peaks in the FFT indicates that a large number of sloshing modes will be required to represent the sloshing loads obtained from CFD. However, the loads cannot be accurately captured by the EMS model since its inherent linear nature. Hence, from the above analysis it can be concluded that the sloshing loads obtained from CFD and EMS cannot be directly compared and the difference in the flutter boundary of the wing section using CFD and EMS model for sloshing can be explained. Although the limitations of EMS model is apparent from the last study, it still remains a useful and computationally inexpensive model for sloshing.

**F. Computational Cost Analysis of the High-Fidelity CFD Solutions and RBF-NN based Surrogate Model**

The computational cost is measured in terms of CPU time of simulation running on an *Intel® Xeon® CPU E5-1650 V3 @ 3.50 GHz* processor running on single core and is summarized in Table 1. The computation times of CFD and RBF-NN are compared for computation/prediction of sloshing loads when the partially filled tank is excited by the same structural motion. The computation time for generation of training data and training the RBF-NN is not considered for comparison.

Table 1. Computational Cost Analysis of High Fidelity CFD and Surrogate Predictive Model

|  | Time (in seconds) | Total Time (in seconds) |
|---|---|---|
| **CFD Simulations** |  |  |
| Sloshing (100 time steps) | 289.56 | 289.56 |
| **Surrogate Predictive Model** |  |  |
| Training Data (900 time steps) | 2382.48 | 2542.68 |
| Training Time | 160.20 |  |
| Sloshing (100 time steps) | 14.51 | 14.51 |
| **Computational Savings** | (289.56-14.51) / 289.56*100 = **94.98%** |  |

The computational savings obtained by the RBF-NN based surrogate model justifies its development and usage for the flutter problem. The model requires one-shot training with the training samples data and can be re-used for prediction and analysis as long as the wing-section motion amplitude and frequency remains in the proximity range of the training samples utilized. The high computational savings is a motivation for more efficient training of the RBF-NN for a reduction in prediction error and robustness to flow conditions and aeroelastic motion.

## V. Conclusions

This work is an attempt to address the lack of computational tools to study the effects of fuel sloshing on aeroelastic characteristics of airfoil and wing in transonic flight. Computational frameworks comprising of various models for sloshing, including high-fidelity CFD model, data-driven surrogate model and the incorporation of a linear Electromechanical System (EMS) model accounting for the effects of sloshing into the structural model are explored. The flutter boundary of a NACA64A010 wing section is computed in the transonic flow range with and without the effects of sloshing in an embedded fuel tank using high-fidelity CFD solutions in time-domain. A change in flutter onset is observed with a general trend of higher flutter speed index with sloshing effects.

The flutter boundary of the same wing section computed with the effects of fuel sloshing computed by high-fidelity CFD and RBF-NN based surrogate model is compared. A remarkable agreement in the flutter onset is observed for both models. The computational savings obtained from the surrogate model for fuel sloshing is computed to be about 95%, which justifies the development of the surrogate model. The data-driven surrogate model is a black-box solution that emulates the sloshing behavior based upon CFD data and operates in a confined range determined by the training samples. Although the sloshing loads prediction by the surrogate model is extremely accurate for the present case, the same prediction quality cannot be guaranteed for other flow regimes.

However, the sloshing surrogate model can be tuned for more accurate predictions and made more robust to external flow conditions by using more rich training data. The present study indicates that even if the prediction accuracy is improved at the expense of computational savings, there is enough leeway for the computational benefits of the surrogate model. This study has demonstrated promising results motivating the feasibility of an extension to a three-dimensional wing with an internal fuel tank.

The flutter boundary of the same wing section computed using EMS model for sloshing is compared with that obtained from CFD and RBF-NN models. The correlation between sloshing effects on the wing section flutter boundary predicted based on the EMS model and the CFD model is poor. This is attributed to the inability of the EMS model to emulate the system nonlinearities of violent sloshing in the fuel tank. A parametric study of frequency response of sloshing using CFD shows the presence of nonlinearities at higher frequencies of excitation, which cannot be effectively captured by the EMS model. However, the EMS formulation is still a reliable and computationally inexpensive tool for modeling sloshing in the linear regime. The various models for sloshing considered in this work support iterative design optimization and flutter mitigation problems at significantly lower computational costs. The prediction accuracy and limitations of data-driven surrogate model and simplified linear model for sloshing has been reviewed in the present work.